\documentclass[lettersize,journal]{IEEEtran}
\usepackage{amsmath,amsfonts}
\usepackage{algorithmic}
\usepackage{array}
\usepackage[caption=false,font=normalsize,labelfont=sf,textfont=sf]{subfig}
\usepackage{textcomp}
\usepackage{stfloats}
\usepackage{url}
\usepackage{verbatim}
\usepackage{graphicx}
\def\BibTeX{{\rm B\kern-.05em{\sc i\kern-.025em b}\kern-.08em
    T\kern-.1667em\lower.7ex\hbox{E}\kern-.125emX}}
\usepackage{balance}

\usepackage[nolist,nohyperlinks]{acronym}

\usepackage{moreverb,url}
\usepackage[nolist,nohyperlinks]{acronym}

\usepackage[colorlinks,bookmarksopen,bookmarksnumbered,citecolor=red,urlcolor=red]{hyperref}

\usepackage{lipsum,array}
\newcolumntype{P}[1]{>{\centering\arraybackslash}p{#1}}
\newcolumntype{C}[1]{>{\centering\arraybackslash}m{#1}}

\usepackage{hyperref}

\usepackage{enumitem}

\newcommand\scalemath[2]{\scalebox{#1}{\mbox{\ensuremath{\displaystyle #2}}}}

\begin{document}
\title{Unified incremental nonlinear controller for the transition control of a hybrid dual-axis tilting rotor quad-plane}
\author{Alessandro Mancinelli, Bart D.W. Remes, Guido C.H.E. de Croon and Ewoud J.J. Smeur
\thanks{This work was carried out within the Unmanned Valley Project. The authors would like to thank the "Europees Fonds voor Regionale Ontwikkeling(EFRO)" who is founding the Unmanned Valley project under grant code KvW-00168 for the South-Holland region.}}


\begin{acronym}
\acro{CNC}{Computerised Numerical Control}
\acro{IMU}{Inertial Measurement Unit}
\acro{PWM}{Pulse Width Modulation}
\acro{INDI}{Incremental Non-linear Dynamic Inversion}
\acro{UAV}{Unmanned Aerial Vehicle}
\acro{VTOL}{Vertical Takeoff and Landing}
\acro{TRUAV}{Tilt Rotor Unmanned Aerial Vehicle}
\acro{DOF}{Degrees Of Freedom}
\acro{MPC}{Model Predictive Control}
\acro{LQR}{Linear Quadratic Regulator}
\acro{CA}{Control Allocation}
\acro{SQP}{Sequential Quadratic Programming}
\acro{AoA}{Angle of Attack}
\end{acronym}

\maketitle

\begin{abstract}
Hybrid overactuated \acp{TRUAV} are a category of versatile UAVs known for their exceptional wind resistance capabilities. However, their extensive operational range, combined with thrust vectoring capabilities, presents complex control challenges due to non-affine dynamics and the necessity to coordinate lift and thrust for controlling accelerations at varying airspeeds. Traditionally, these vehicles rely on switched logic controllers with two or more intermediate states to control transitions.
In this study, we introduce an innovative, unified incremental nonlinear controller designed to seamlessly control an overactuated dual-axis tilting rotor quad-plane throughout its entire flight envelope. Our controller is based on an incremental nonlinear control allocation algorithm to simultaneously generate pitch and roll commands, along with physical actuator commands. The control allocation problem is solved using a Sequential Quadratic Programming (SQP) iterative optimization algorithm making it well-suited for the nonlinear actuator effectiveness typical of thrust vectoring vehicles.
The controller's design integrates desired roll and pitch angle inputs. These desired attitude angles are managed by the controller and then conveyed to the vehicle during slow airspeed phases, when the vehicle maintains its 6 Degrees of Freedom (DOF). As the airspeed increases, the controller seamlessly shifts its focus to generating attitude commands for lift production, consequently smoothly disregarding the desired roll and pitch angles.
Furthermore, our controller integrates an Angle Of Attack (AOA) protection logic to mitigate wing stalling risks during transitions. It also features a yaw rate reference model to enable coordinated turns and minimize side-slip. The effectiveness of our proposed control technique has been confirmed through comprehensive flight tests. These tests demonstrated the successful transition from hovering flight to forward flight, the attainment of desired vertical and lateral accelerations, and the ability to revert to hovering.

\end{abstract}

\begin{IEEEkeywords}
UAV, VTOL, Control Allocation, Nonlinear programming, INDI, Tilt rotor, Quad-plane, Fully actuated vehicles, Hybrid MAV, Weighted Least Squares, Quadratic Programming.
\end{IEEEkeywords}

\section{Introduction}

\begin{figure*}
	\centering
	\includegraphics[scale  = .55]{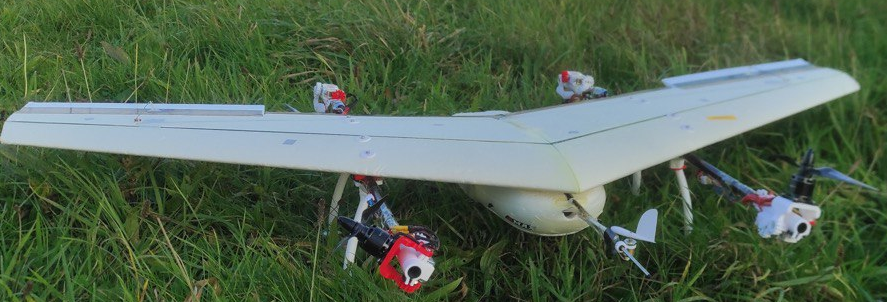}
	\caption{A picture of the hybrid dual-axis tilting rotor quad-plane.}
	\label{Prototype_figure}
\end{figure*}

\acresetall
Hybrid \acp{UAV} represent a unique class of aircraft, seamlessly integrating \ac{VTOL} capabilities typically associated with helicopters with the extended endurance and operational efficiency characteristic of fixed-wing planes. Categories of hybrid lift UAVs that have received significant attention in research include tailsitters \cite{10.2514/1.G004520}, quad-planes \cite{10.1016/j.ast.2021.107105}, tilting wings \cite{10.3390/aerospace5010017}, and tilting rotors \cite{10.1109/ICUAS.2013.6564761}.

Each of these vehicles has its own set of advantages and disadvantages. However, among this diverse spectrum of hybrid \ac{UAV} configurations, one particular type of vehicle has demonstrated exceptional suitability for operations in gusty environments: the \ac{TRUAV}.  \acp{TRUAV} distinguish themselves by their ability to rapidly adjust rotor orientation, enabling them to respond swiftly to wind disturbances and minimize the impact of wind on the vehicle's position \cite{10.1109/ICUAS54217.2022.9836063}. Additionally, \acp{TRUAV} can be designed as overactuated systems, allowing for independent control of both their position and attitude. This characteristic enhances their versatility and adaptability \cite{10.3182/20140824-6-ZA-1003.02692}.

Nonetheless, \acp{TRUAV} encounter relevant control challenges that have somewhat constrained their widespread adoption. The primary challenge in controlling \acp{TRUAV} lies in their control/input-non affine dynamics, which presents a significant hurdle. Control/input-non affine dynamics implies that the system exhibits non-linearities both in its state and control inputs. Unlike conventional \acp{UAV}, where thrust can be assumed to be consistently applied in the same direction in the body frame, \acp{TRUAV} have the capacity to alter the direction of thrusters within the body frame. This characteristic results in a pronounced nonlinear relationship between the control inputs and the generated vehicle accelerations. Consequently, when the tilting dynamics are particularly rapid, the \ac{CA} problem becomes increasingly complex and cannot be adequately addressed using conventional algorithms relying on linearized control effectiveness simplifications \cite{10.1007/s10846-023-01865-8}.

Due to the inherent complexity of the system, the number of published works addressing the control of \ac{TRUAV} remains limited. One such notable work was conducted by Papachristos et al. \cite{10.1109/ICRA.2013.6631355}. In their work, they tackled the control/input-non affine dynamics challenge of rotor tilting by implementing a hybrid \ac{MPC}/PID controller. They utilized a nonlinear \ac{MPC} to generate tilting and attitude commands. Subsequently, the tilting commands were sent to the actuators, while the attitude commands were executed through a gain-scheduled PID controller controlling the motor commands. In a more recent study \cite{10.1007/s10846-023-01865-8}, we developed a nonlinear incremental \ac{CA} algorithm employing a nonlinear solver capable of directly controlling both linear and angular accelerations through the generation of motor and tilt commands. However, it is important to note that both of these works primarily focused on controlling the vehicle in the hovering configuration, with limited analysis of the transition to forward flight. Consequently, this limitation significantly restricts the applicability of these controllers, mainly to low airspeeds.

The reason why these two works have been primarily restricted to low airspeed is due to a significant challenge encountered by hybrid \acp{TRUAV} during the transition from low-speed to high-speed flight regimes. This transition necessitates efficient allocation of desired accelerations between direct actuator commands and the vehicle's attitude. To illustrate this challenge, let's consider the transition control from low-speed to high-speed flight of the dual-axis tilt rotor quad-plane, presented in \cite{10.1109/ICUAS54217.2022.9836063} and depicted in Figure \ref{Prototype_figure}.

At low airspeed, the vehicle's 4 motors and 8 tilting servos provide direct control over all 6 \ac{DOF}, enabling the achievement of any desired linear and lateral acceleration through direct actuator commands. However, as the vehicle transitions to high airspeed, achieving vertical acceleration becomes more efficiently managed through pitch adjustments rather than motor or tilt commands. Additionally, lateral acceleration can only be attained through roll adjustments, as the lateral tilt actuator becomes ineffective due to rotor gimbal lock. The gimbal lock condition of the lateral rotor tilt can be understood by referring to Figure \ref{fig:rotor_angle_definition}, where the tilting angles of the dual-axis tilting rotor quadplane are defined. Assuming the rotors are fully oriented forward (e.g., with the tilt angle $b$ set at -90 degrees), a lateral tilt angle rotation $g$ results in a rotation around the spinning axis of the motor, rendering it ineffective for tilting the thrust. Consequently, despite the vehicle possessing 6 \ac{DOF} in the hovering configuration, it retains only 4 \ac{DOF} authority in the high-speed regime. This necessitates the development of a distinct control strategy to achieve lateral and vertical accelerations in different flight regimes.

The issue of re-allocating the accelerations over different actuators or states at different flight regimes is well known in the literature. The most common way to overcome this issue for this class of vehicles has been to separate the vehicle's transition control using one or multiple intermediate states. Those states are then used together with a switched logic controller to control each flying phase separately \cite{10.1016/j.ast.2015.10.012, 10.2514/1.G000263}. The controller developed at each intermediate step to control the vehicle ranges from \ac{LQR} \cite{10.5281/zenodo.1074431} to simple PID \cite{10.1016/j.mechatronics.2012.03.003}. The main disadvantage of these controllers is their lack of continuous control, the limited performance optimization and disturbance rejection capability.

A more advanced controller was introduced by Raab et al. \cite{10.2514/6.2018-3478}, who developed a unified control framework for \ac {TRUAV} vehicles. This proposed control solution employs a single-loop \ac{INDI} control structure. At the core of their incremental control structure is the generation of pitch and roll commands directly within the \ac{CA} inversion loop, referred to as virtual commands. These virtual commands are derived from the linearized effectiveness of pitch and roll angles and then applied to the system using an \ac{INDI} control scheme. 

While this control architecture shows promise and has yielded favorable results, it does have certain limitations. Notably, the proposed control architecture does not include the tilting angle as a control input. By doing so, the authors have reasonably assumed the control input dynamics as affine and used linear approximation for control effectiveness to solve the \ac{CA} problem. Consequently, this controller may not be suitable for vehicles equipped with quick-tilting actuators that can be directly employed into the vehicle's control. Furthermore, the \ac{CA} algorithm employed in this framework does not facilitate independent attitude and position control. This capability is particularly crucial for precision landing on a tilting platform, such as a moving vessel.

In this work, we propose a unified nonlinear control framework for a hybrid dual-axis tilting rotor quad-plane capable of addressing all the previously analyzed \ac{TRUAV} control challenges.
Taking inspiration from the virtual commands approach proposed in \cite{10.2514/6.2018-3478}, we modified the cost function of the nonlinear optimization process presented in \cite{10.1007/s10846-023-01865-8} in such a way to compute the references for the roll and the pitch angle inside the nonlinear \ac{CA} algorithm.
Additionally, we implemented further adjustments in the error control to ensure smooth lateral acceleration tracking across all flight regimes. We also integrated an Angle of Attack (AoA) protection system into the nonlinear optimization process and we have incorporated desired pitch and roll commands into the secondary objective of the cost function.
As a result of these enhancements, the vehicle can achieve full 6 \ac{DOF} authority in the low-speed regime while smoothly discarding external desired pitch and roll values as airspeed increases. As a final development, an accelerometer based sideslip controller was included in the yaw rate reference model, enabling coordinated turns and minimizing the sideslip angle.

The developed control strategy has been implemented and tested on the dual-axis tilting rotor quad-plane depicted in Figure \ref{Prototype_figure}. Real-time computation for the 13 actuators and the two virtual attitude commands was performed using a Raspberry Pi 4B companion computer, at an average refresh rate of 220 Hz.

Results from flight tests have demonstrated the effectiveness of the control strategy. Throughout the flight test maneuvers, the vehicle exhibited precise tracking of linear and angular acceleration and effectively managed the loss of \ac{DOF} during the transition.

The paper is structured as follows: in section \ref{sec:Nonlinear_CA_algorithm_derivation}, we derive the incremental Nonlinear Unified \ac{CA} problem which generates the commands for both the attitude and the physical vehicle actuators. Section \ref{sec:Unified_Incremental_Nonlinear_Control_Framework} introduces the control framework that supports the Incremental Nonlinear Unified \ac{CA} algorithm for controlling the dual-axis tilt rotor quad-plane. In section \ref{sec:Controller_setup_and_parameters}, we conclude the setup of the unified controller providing also some characteristic parameters of the flying vehicle. Section \ref{sec:Flight_test_experiment_and_results} is dedicated to presenting and analyzing the results of the flight tests. In section \ref{sec:Limitations_of_the_Unified_Incremental_Nonlinear_Controller}, we discuss some limitations of the Nonlinear Unified \ac{CA} algorithm. Finally, section \ref{sec:Conclusions} summarizes our conclusions.

\section{Nonlinear Control Allocation problem definition}
\label{sec:Nonlinear_CA_algorithm_derivation}
In this section, we will formulate the Unified Incremental Nonlinear Control Allocation problem for the vehicle. Initially, we will derive the generic Nonlinear Control Allocation formulation, then we will extend the algorithm to accommodate the virtual actuators of pitch and roll angle.

Considering the nonlinear system characterizing the dynamics of our flying vehicle:

\begin{equation}
\left\{ \begin{array}{c}
    \boldsymbol{\dot{x}}(t) = h(\boldsymbol{x}(t),\boldsymbol{u}(t)) \\
    \boldsymbol{y}(t) = g(\boldsymbol{x}(t)) \\
\end{array} , \right.
\end{equation}
where, $ \boldsymbol{x}(t) \in \mathbb{R}^n$ represents the state vector, $ \boldsymbol{u}(t) \in \mathbb{R}^m$ is the control input vector, and $ \boldsymbol{y}(t) \in \mathbb{R}^p$ is the output vector. The function $h : D_{\boldsymbol{x},\boldsymbol{u}} \rightarrow \mathbb{R}^n$ represents the nonlinear state dynamics of the vehicle while the function $g : D_{\boldsymbol{x}} \rightarrow \mathbb{R}^p$ is the nonlinear output dynamics. Let's now assume the output vector $\boldsymbol{y}$ to be composed of the linear and angular speed components of the vehicle: 
\begin{equation*}
\boldsymbol{y} = \left(\dot{x}, \; \dot{y}, \; \dot{z}, \; p, \; q, \; r \right).
\end{equation*}
We can then define a nonlinear function $f : D_{\boldsymbol{x},\boldsymbol{u}} \rightarrow \mathbb{R}^p$ characterizing the nonlinear dynamics of the output vector derivative $\boldsymbol{\dot{y}}$ as a function of the vehicle state $\boldsymbol{x}$ and control input $\boldsymbol{u}$:
\begin{equation}
    \boldsymbol{\dot{y}}(t) = f(\boldsymbol{x}(t),\boldsymbol{u}(t)).
    \label{system_description}
\end{equation}

Starting from Equation \ref{system_description}, we can derive the acceleration increment of the system:

\begin{equation}
\begin{aligned}
&\boldsymbol{\dot{y}}(t_0 + \Delta t) - \boldsymbol{\dot{y}}_0 = \\
= f(\boldsymbol{x}(t_0 + &\Delta t),\boldsymbol{u}(t_0 + \Delta t)) - f(\boldsymbol{x}_0,\boldsymbol{u}_0).
\end{aligned}
\label{eq:indi_increment_1}
\end{equation}
Where $\boldsymbol{\dot{y}}_0$, $\boldsymbol{u}_0$ and $\boldsymbol{x}_0$ represent the current vehicle accelerations, control input and state vector at time $t=t_0$.

Under the assumption of a large actuator bandwidth and small time increment $\Delta t$, we can consider that the changes in acceleration primarily depend on the system inputs. Recent research has shown the validity of this assumption for linearized systems \cite{10.2514/1.G007079}. Here, we extend this assumption to encompass nonlinear systems, simplifying Equation \ref{eq:indi_increment_1} to:

\begin{equation}
\boldsymbol{\dot{y}}(t_0 + \Delta t) - \boldsymbol{\dot{y}}_0 = f(\boldsymbol{x}_0,\boldsymbol{u}(t_0 + \Delta t)) - f(\boldsymbol{x}_0,\boldsymbol{u}_0).
\label{eq:indi_increment_2}
\end{equation}

From equation \ref{eq:indi_increment_2}, we can now identify a control input command $\boldsymbol{u}_{cmd}$ at time $t_0 + \Delta t$ needed to reach desired accelerations $\boldsymbol{\dot{y}}_d$ at the same time step:
\begin{equation}
\begin{aligned}
f( &\boldsymbol{x}_0,\boldsymbol{u}_{cmd}(t_0 + \Delta t)) = \\
= f(\boldsymbol{x}_0,&\boldsymbol{u}_0) - \boldsymbol{\dot{y}}_0 + \boldsymbol{\dot{y}_d}(t_0 + \Delta t).
\end{aligned}
\label{eq:indi_increment_3}
\end{equation}

Unfortunately, the control input $\boldsymbol{u}_{cmd}$ is still embedded within a complex nonlinear function. To find its value, we need to use a nonlinear optimization algorithm. To do this, we can formulate a nonlinear cost function to minimize in order to determine the value of $\boldsymbol{u}_{cmd}$:
\begin{equation}
\scalemath{0.95}{
\begin{aligned}
&C_p(\boldsymbol{u}_{cmd}(t_0 + \Delta t)) = \\
= ||\boldsymbol{W_\nu} \left[f(\boldsymbol{x}_0,\boldsymbol{u}_{cmd}(\right. &t_0 + \Delta t)) + \boldsymbol{\dot{y}}_0 - f(\boldsymbol{x}_0,\boldsymbol{u}_0) + \\ 
&\left. - \boldsymbol{\dot{y}}_d(t_0 + \Delta t)\right] ||^2.
\end{aligned}
}
\label{eq:indi_increment_4}
\end{equation}

Where the current linear and angular accelerations of the vehicle $\boldsymbol{\dot{y}}_0$ can be measured using accelerometers and gyros. The diagonal matrix $\boldsymbol{W_\nu}$ serves as a weighting matrix for the desired accelerations. The elements within this matrix allow for the prioritization of specific acceleration increments over others. A larger value in the matrix element results in a higher priority for the associated acceleration increment. It's important to note that the $\boldsymbol{u}_{cmd}$, determined through the minimization of the cost function in equation \ref{eq:indi_increment_4}, represents the output of an incremental law since it is calculated based on an acceleration increment.

Now, we can integrate the cost function $C_p$ from equation \ref{eq:indi_increment_4} into a larger cost function $C$, which also includes a secondary objective aimed at minimizing the overall control input magnitude. This additional term is particularly significant for over-actuated vehicles, where the control input solution associated with a desired acceleration array $\boldsymbol{\dot{y}}_d$ may not be unique. The overall cost function $C$ is then: 
\begin{equation}
\begin{aligned}
    &C(\boldsymbol{u}_{cmd}(t_0 + \Delta t)) = C_p(\boldsymbol{u}_{cmd}(t_0 + \Delta t)) + \\
    &+ ||\gamma_u^{\frac{1}{2}}\boldsymbol{W_u} (\boldsymbol{u}_{cmd}(t_0 + \Delta t) - \boldsymbol{u_d} (t_0 + \Delta t) )||^2 .   
\end{aligned}
\label{eq:cost_function_final}
\end{equation}

In this context, $\boldsymbol{u_d}(t_0 + \Delta t)$ represents the desired control input vector at time $t_0 + \Delta t$, and $\gamma_u$ is the control input optimality scale factor. It is crucial to maintain a small value for $\gamma_u$ ($\gamma_u << 1$) as it serves to differentiate the primary objective from the secondary objective within the cost function.

Regarding the diagonal matrix $\boldsymbol{W_u}$, it acts as the control input weighting matrix. The diagonal elements of this matrix assign priority to the usage of specific actuators over others. The larger the diagonal element of the control input weighting matrix, the higher the cost associated with that particular actuator. It's important to note that the cost of each actuator is always measured in relation to the desired control input value $\boldsymbol{u_d}(t_0 + \Delta t)$.

Once that the cost function is derived, we can define the constrained \ac{CA} optimization problem associated with the cost function in equation \ref{eq:cost_function_final}:
\begin{equation}
\scalemath{1}{
\begin{gathered}
     \boldsymbol{u_s} = \text{arg  min } C(\boldsymbol{u}_{cmd}) \\
     \text{subject  to} \\
     \boldsymbol{u}_{min} < \boldsymbol{u} < \boldsymbol{u}_{max}.
\end{gathered}
}
\label{Nonlinear_CA_method}
\end{equation}

Where $C(\boldsymbol{u}_{cmd})$ represents the nonlinear cost function as defined in equation \ref{eq:cost_function_final}, with $\boldsymbol{u}_{max}$ and $\boldsymbol{u}_{min}$ specifying the constraints on the control input.
In order to reduce problem complexity, the \ac{CA} problem associated with our vehicle employs a simplified version of the system dynamics $\boldsymbol{f}_s(\boldsymbol{x}, \boldsymbol{u})$ instead of the full system dynamics $\boldsymbol{f}(\boldsymbol{x}, \boldsymbol{u})$. A comprehensive derivation of the simplified system dynamics for the dual-axis tilting rotor quad-plane is provided in the Appendix.

Having defined the \ac{CA} problem in equation \ref{Nonlinear_CA_method}, the next step involves the normalization of this problem. The normalization process is crucial to enable the optimization algorithm to equally consider actuators with different units of measure and travel.

The normalization process involves scaling all the terms of the cost function in Equation \ref{Nonlinear_CA_method}, in a way that the cost function can be evaluated using a normalized control input vector $\boldsymbol{u}^*$ instead of $\boldsymbol{u}$.
The normalized input vector $\boldsymbol{u}^*$ is defined as follows:
\begin{equation}
     \boldsymbol{u}^* = \frac{\boldsymbol{u}}{\boldsymbol{G}},
\label{control_input_array_scaled}
\end{equation}
where 
\begin{equation}
\boldsymbol{G} = \frac{\boldsymbol{u}_{max} - \boldsymbol{u}_{min}}{2}.
\label{eq:scaling_array_G}
\end{equation}
In the normalized input vector $\boldsymbol{u}^*$, all actuators have a travel equal to 2 units, regardless of their physical travel. The optimization process will provide a normalized actuator solution as output, which must be re-scaled.

The normalized Unified Nonlinear Control Allocation problem then becomes:
\begin{equation}
\scalemath{1}{
\begin{gathered}
     \boldsymbol{u_s^*} = \text{arg  min } C^*(\boldsymbol{u}^*_{cmd}) \\
     \text{subject  to} \\
     \boldsymbol{u}_{min}^* < \boldsymbol{u^*} < \boldsymbol{u}_{max}^*
\end{gathered}
}
\label{Normalized_control_problem}
\end{equation}
where
\begin{equation*}
\begin{aligned}
    &C^*(\boldsymbol{u}^*_{cmd}) = C_p(\boldsymbol{u}^*_{cmd} \cdot \boldsymbol{G}) + \\
    &+ ||\gamma_u^{\frac{1}{2}}\boldsymbol{W_u} (\boldsymbol{u}^*_{cmd} - \boldsymbol{u}_d/\boldsymbol{G} )||^2 .   
\end{aligned}
\end{equation*}
and
\begin{equation*}
\scalemath{1}{
   \boldsymbol{u}_s = \boldsymbol{u}_s^* \cdot \boldsymbol{G},
   }
\end{equation*}
Here, $\boldsymbol{G}$ represents the scaling vector defined in Equation \ref{eq:scaling_array_G}, and $\boldsymbol{u}_s$ denotes the re-scaled control input solution. All products and divisions involving the scaling vector $\boldsymbol{G}$ are performed element-wise.

\subsection{Control Input vector definition}
\begin{figure}
	\centering
	\includegraphics[scale  = .21]{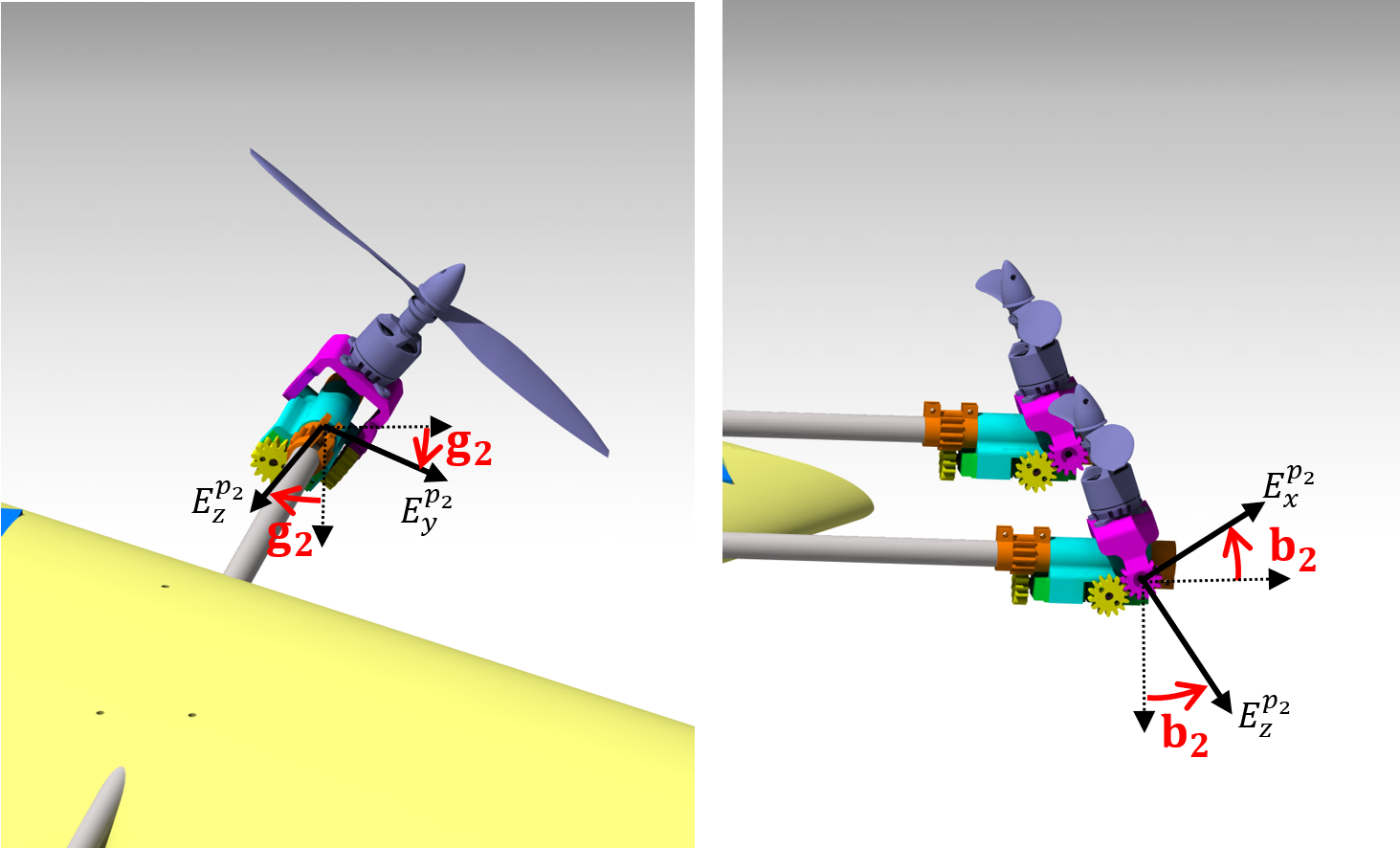}
	\caption{Definition of the rotor elevation angle $b$ and azimuth angle $g$.}
	\label{fig:rotor_angle_definition}
\end{figure}

\begin{figure}
	\centering
	\includegraphics[scale  = .4]{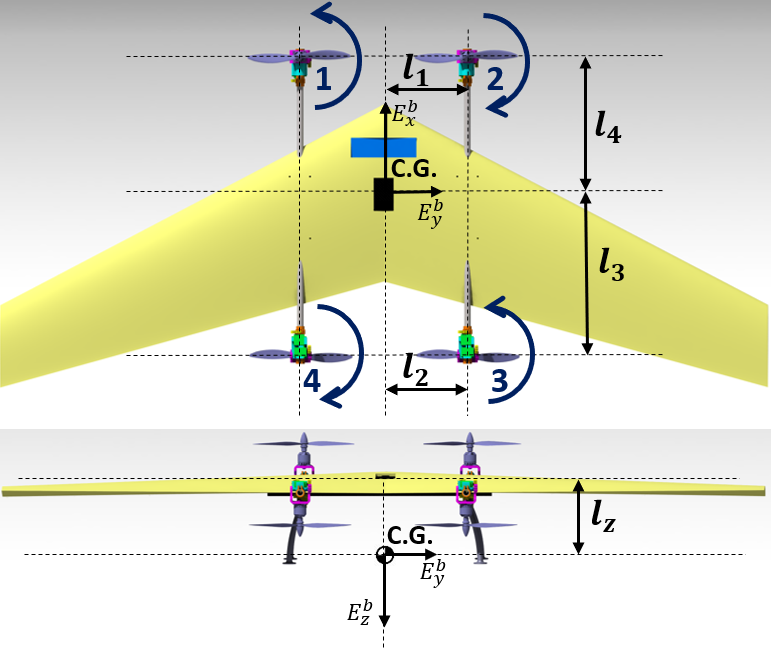}
	\caption{Motor spinning direction, rotor numbering and vehicle geometry with respect to the Center of Gravity (CG).}
	\label{fig:Assumptions_and_notation}
\end{figure}

Regarding the definition of the control input vector $\boldsymbol{u}$ utilized in the Control Allocation Optimization problem, we have adopted the subsequent control input vector:
\begin{equation}
\boldsymbol{u} = ( \boldsymbol{u}_{a} , \phi_v , \theta_v )^T.
\label{control_input_vector}
\end{equation}
Here, the first element, $\boldsymbol{u}_{a}$, represents an array containing the physical command to each of the actuators, while the last two elements, $\phi_v$ and $\theta_v$, are referred to as virtual commands. 
The $\boldsymbol{u}_{a}$ vector for the dual-axis hybrid quad-plane is composed of the following elements: 
\begin{equation}
\begin{aligned}
   \boldsymbol{u}_{a} = \; & \left( \; \Omega_1 \; \Omega_2 \; \Omega_3 \; \Omega_4 \right. \\
    & \hspace{.5cm} b_1 \; b_2 \; b_3 \; b_4 \; \\
    & \left. \hspace{.2cm} g_1 \; g_2 \; g_3 \; g_4 \; \delta_a \right), \\
   \end{aligned}
\label{actuators_list}
\end{equation}
where:
\begin{itemize}
    \item $\Omega_i$ is the rotational speed of the i-th rotor.
    \item $b_i$ is the longitudinal tilting angle of the i-th rotor (elevation tilting angle). 
    \item $g_i$ is the lateral tilting angle of the i-th rotor (azimuth tilting angle). 
    \item $\delta_a$ is the aileron deflection angle, positive for a positive generation of roll rate. 
\end{itemize}
The rotor deflection angles are defined in Figure \ref{fig:rotor_angle_definition} while the rotor numbering definition is displayed in Figure \ref{fig:Assumptions_and_notation}. 

The unconventional definition of the control input vector in Equation \ref{control_input_vector}, encompassing both physical and virtual actuators, was essential to devise a control law capable of seamlessly allocating the desired acceleration across varying airspeed conditions.

In the hovering configuration, at low airspeed, the vehicle's actuators can directly generate all three forces and three moments required for the 6 \ac{DOF} control of the vehicle. However, as the airspeed increases, we need to incorporate lift as a means of controlling lateral and vertical acceleration.

To accomplish this, we incorporated roll and pitch angles as active optimization control inputs. These angles are no longer treated as current vehicle states but are considered as variables to be optimized, similar to the physical actuators of the system. Leveraging knowledge of the vehicle's aerodynamic properties, the optimization process determines whether it's more efficient to achieve a certain linear acceleration by actuating the physical actuators or by adjusting the vehicle's attitude. Fine-tuning of the controller prioritization of attitude commands over physical actuators can be achieved acting on the control input weighting matrix $\boldsymbol{W_u}$, which in our case incorporates a dependency on airspeed.

In our specific case, we have selected the control input weighting matrix $\boldsymbol{W_u}$ with the intention of discouraging the utilization of lateral tilt angles at high airspeed, while simultaneously promoting the utilization of roll and pitch angles as airspeed increases:
\begin{equation}
\begin{aligned}
    \boldsymbol{W_u} = & diag \left( 3 \;,\; 3 \;,\; 3 \;,\; 3 \;,\; 0 \;,\; 0 \;,\; 0 \;,\; 0 \right. \\
                \\[-1.2em]
                &  1.5 V_{a} \;,\;  1.5 V_{a} \;,\;  1.5 V_{a} \;,\;  1.5 V_{a} \\
                \\[-1.2em]
                &  \left. 0.5 \;,\; 100 - 15 V_{a_b} \;,\; 100 - 15 V_{a_b} \right).
\end{aligned}   
\end{equation}
Where $V_{a}$ represents the airspeed measured by the Pitot tube in m/s. Note that the last two diagonal elements of the $\boldsymbol{W_u}$ use a different definition of the airspeed, named $V_{a_b}$. $V_{a_b}$ is a bounded airspeed value within 0 and 100/15 m/s, to avoid the elements of $\boldsymbol{W_u}$ to become negative.

Including the generation of the roll and pitch commands in the control input array offers an additional advantage. It allows for the incorporation of desired roll and pitch commands in the desired actuator command array $\boldsymbol{u}_d$, which forms part of the secondary objective in the cost function. This implies that the vehicle will continually try to attain the desired roll and pitch condition specified in the $\boldsymbol{u}_d$ array.

However, as airspeed increases, the vehicle will increasingly rely on roll or pitch changes to achieve the desired lateral or vertical acceleration increment. As acceleration increments are a component of the primary objective within the cost function, the roll and pitch commands will be primarily computed to fulfill lateral and vertical accelerations. Consequently, as airspeed increases, the desired roll and pitch values in $\boldsymbol{u}_d$ will be gradually and automatically disregarded.

\subsection{Accelerations weighting matrix \texorpdfstring{$\boldsymbol{W_{\nu}}$}{Lg} and optimality scale factor \texorpdfstring{$\gamma_u$}{Lg}}
To complete the Nonlinear Control Allocation derivation, we need to define the optimality scale factor $\gamma_u$ and the desired accelerations weighting matrix $\boldsymbol{W_{\nu}}$.

The optimality scale factor $\gamma_u$, as explained in the previous subsection, separates the primary objective from the secondary objective of the cost function. After extensive testing, a value of $\gamma_u = 1e^{-6}$ was selected for the vehicle.

Regarding the desired accelerations weighting matrix $\boldsymbol{W_\nu}$, we chose to prioritize attitude commands first, followed by altitude commands, and finally lateral and longitudinal commands. The desired accelerations weighting matrix $\boldsymbol{W_\nu}$ selected for the vehicle is as follows:
\begin{equation}
\scalemath{1.0}{
    \boldsymbol{W}_\nu = \text{diag }\left( .005, \; .005, \; .008, \; .015, \; .015, \; .015 \right).
    }
\end{equation}
It is considered a best practice to give higher priority to attitude control over linear accelerations. This practice becomes even more significant in the context of traditional, non-thrust vectoring vehicles. In these cases, the direction of thrust can only be effectively applied when the vehicle maintains the correct attitude.

\section{Unified Incremental Nonlinear Control Framework}
\label{sec:Unified_Incremental_Nonlinear_Control_Framework}
In this section, we present the unified incremental nonlinear control framework, which consists of three main blocks.

\begin{figure*}
	\centering
	\includegraphics[scale  = .95,trim={.8cm .6cm 1.2cm .8cm},clip]{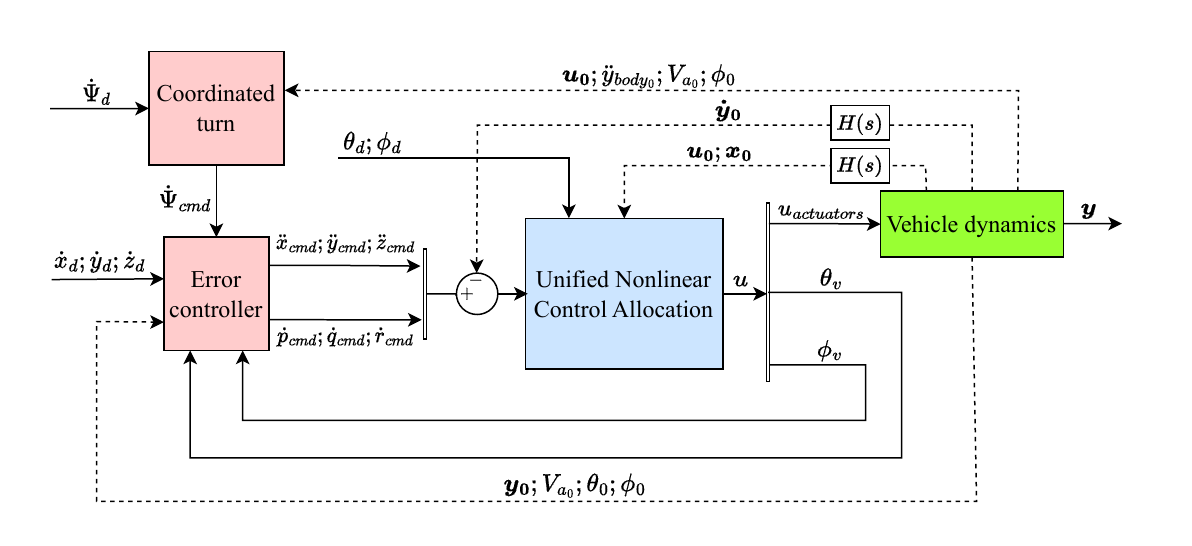}
	\caption{Diagram of the Unified Incremental Nonlinear Controller. In the diagram, the dashed lines represent variables estimated through sensor (for the vehicle states and accelerations) or actuator model (for the current actuator state vector). The red blocks are running on the primary flight computer, the blue block runs on the Raspberry pi while the green block represents the evolution of the vehicle dynamics.}
	\label{Nonlinear_unified_controller_scheme}
\end{figure*}

\begin{enumerate}[left=0pt, itemsep=0em]

    \item \textbf{Coordinated Turn Block:} This block is dedicated to generating the yaw rate reference. Its role is to control turns during high-airspeed flight and minimize the sideslip angle of the vehicle.
    
    \item \textbf{Error Controller block:} This block is responsible for generating linear and angular acceleration commands. These commands are subsequently relayed to the Unified Nonlinear Control Allocation block.
    
    \item \textbf{Unified Nonlinear Control Allocation Block:} This block takes as input the current vehicle states and desired acceleration increments. It then generates commands for both the physical actuators and the desired pitch and roll angles.
    
\end{enumerate}

A schematic representation of the unified incremental nonlinear control framework is provided in Figure \ref{Nonlinear_unified_controller_scheme}. Further elaboration on each block is presented in the subsequent subsections.

\subsection{Coordinated turn block}
The coordinated turn block generates the yaw rate command, which is then fed into the error controller. This yaw rate command plays a crucial role at high airspeed and during transition, as it enables the execution of a coordinated turn while minimizing side-slip. 
The governing equation for coordinated turn block is as follows:
\begin{equation}
     \dot{\Psi}_{cmd} = 9.81 \frac{ \tan(\phi)}{V_{a_c}} K_{\dot{\Psi}_{air}} - \dot{y}_{c}   K_\beta + \dot{\Psi}_d
\label{coordinated_turn_block_law}
\end{equation}
Where $\dot{\Psi}_d$ is the desired hovering yaw rate, $K_\beta$ is the side-slip correction gain, $\dot{y}_{c}$ is the corrected body lateral acceleration and $K_p^T$ is the thrust coefficient of the motor, identified in the next section. The governing equation for the coordinated turn block, as presented in equation \ref{coordinated_turn_block_law}, comprises three primary components.

The first component is a roll angle feed-forward, which is essential for executing coordinated turns effectively, as elaborated in detail in \cite{10.1049/cp.2012.1081}. To prevent undesired yaw actions at low speed, the gain $K_{\dot{\Psi}_{air}}$ was introduced. This gain scales the roll angle feed-forward component, and it is defined as follows:
\begin{equation}
\scalemath{.94}{
K_{\dot{\Psi}_{air}} = 
\begin{cases}
 0  & \hspace{.3cm} V_{a} < V_{\dot{\Psi}_{min}}\\ 
 \noalign{\vskip3pt}
 \cfrac{V_{a} - V_{\dot{\Psi}_{min}}}{V_{\dot{\Psi}_{ref}} - V_{\dot{\Psi}_{min}}}  & 
\hspace{.3cm} V_{\dot{\Psi}_{min}} < V_{a} < V_{\dot{\Psi}_{ref}} \\
 \noalign{\vskip3pt}
 1  & \hspace{.3cm} V_{a} > V_{\dot{\Psi}_{ref}}.
\end{cases}
}
\end{equation}
This gain is crucial because, at low airspeed, the vehicle can achieve the desired roll and pitch angle. When the desired roll angle is attained, the feed-forward component of the control law in equation \ref{coordinated_turn_block_law} may generate an undesirable yaw rate command. The $K_{\dot{\Psi}_{air}}$ gain is designed to reduce the roll feed-forward term to zero below a minimum speed $V_{\dot{\Psi}_{min}}$. The gain gradually increases to 1 between $V_{\dot{\Psi}_{min}}$ and $V_{\dot{\Psi}_{ref}}$ to ensure smooth transitions in the yaw rate command. For the dual-axis tilting rotor quad-plane, the reference speeds for the coordinated turn block are chosen as follows:
\begin{equation}
    V_{\dot{\Psi}_{ref}} = 4 \hspace{.1cm} \frac{m}{s}  \hspace{.5cm},
    V_{\dot{\Psi}_{min}} = 6 \hspace{.1cm} \frac{m}{s}.
\end{equation}
The feed-forward roll angle law component in equation \ref{coordinated_turn_block_law} assumes a consistently positive Angle of Attack (AoA). This assumption implies that a distinct yaw rate law is necessary for scenarios where the vehicle executes turns at a negative AoA.
Furthermore, the term $V_{a_c}$ within equation \ref{coordinated_turn_block_law} represents a corrected airspeed value, defined as follows:
\begin{equation*}
V_{a_c} = \max(10, V_a);
\end{equation*}
This correction is essential to prevent the feed-forward term from reaching infinity when $V_a$ is equal to zero.

The second component of the yaw rate generation law in equation \ref{coordinated_turn_block_law} is the side-slip correction term. This term utilizes feedback from the corrected body lateral acceleration $\dot{y}_{c}$ to mitigate side-slip at high airspeed. The corrected body lateral acceleration $\dot{y}_c$ comprises a sensed body acceleration term $\dot{y}_0^{body}$, which is adjusted to eliminate the acceleration contribution induced by the sideways tilting of the rotors:
\begin{equation}
\begin{aligned}
\dot{y}_{c} = &\; \dot{y}_0^{body} - \frac{K_p^T}{m} \left( \Omega_1^2 \sin(g_1) + \Omega_2^2 \sin(g_2) \right. \\ 
&\left. \hspace{.5cm} \Omega_3^2 \sin(g_3) + \Omega_4^2 \sin(g_4) \right) \\
\end{aligned}
\end{equation}
The assumption of a proportional relation between the side-slip angle and the corrected body lateral acceleration is derived from a detailed analysis carried out by Smeur et al. \cite{10.2514/1.G004520}.

The third and final component of the yaw rate generation law in equation \ref{coordinated_turn_block_law} is a desired yaw rate $\dot{\Psi}_d$. This component can be used to manually input a heading change at low airspeed.

\subsection{Error controller block}
For the generation of linear and angular acceleration commands, a linear error controller is employed. The error controller plays a crucial role in correcting inaccuracies in the model and compensating for external disturbances. This is especially important because we are using a simplified version of the system dynamics. Consequently, the secondary terms of the system dynamics that are not considered in the Control Allocation problem are treated as unmodeled dynamics and managed by the error controller, through the feedback lines.

\begin{figure*}
	\centering
	\includegraphics[scale  = .88,trim={1.5cm .1cm .2cm .1cm},clip]{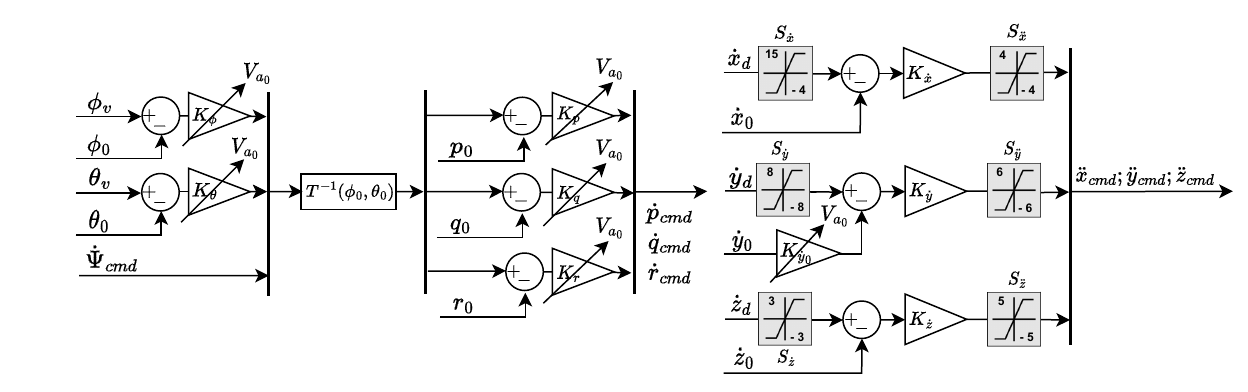}
	\caption{Diagram of the linear Error Controller. The gains value for the dual axis tilt rotor quad-plane are reported in Table \ref{tab:gains}. }
	\label{EC_nonlinear_unified_controller}
\end{figure*}

The error controller consists of variable gains, feedback lines, and saturation blocks, as depicted in Figure \ref{EC_nonlinear_unified_controller}.

The reference frame employed for generating linear accelerations is the control reference frame. This reference frame corresponds to an Earth reference frame, but it is rotated around the Z axis by the vehicle's yaw angle $\psi$. For a visual representation and a comprehensive definition of the control reference frame, reader is referred to the appendix.

Manipulating the linear acceleration and speed terms in this reference frame offers two advantages. Firstly, it simplifies the direct application of speed and acceleration constraints to the aircraft's lateral and longitudinal motion using saturation blocks. This is particularly valuable when imposing maximum forward or lateral speeds during hovering since the components of the control reference frame align with those of the vehicle's body frame, making it possible to directly apply a maximum forward speed using a saturation block. Secondly, it provides the flexibility to eliminate the lateral speed feedback term as airspeed increases. This can be achieved by applying zero gain to the $\dot{y}_0$ feedback line, as illustrated in Figure \ref{EC_nonlinear_unified_controller}.

Disabling the lateral speed feedback line as airspeed increases is essential to prevent the vehicle from accumulating and maintaining a constant lateral speed during high-airspeed flight. This is crucial because, while the vehicle can maintain a constant lateral speed by tilting the rotors sideways at low airspeed, retaining a constant lateral speed in the high-airspeed regime is undesirable and inefficient, as it can lead to side-slip. By deactivating the lateral speed feedback line $\dot{y}_0$, the reference for the desired lateral speed automatically becomes a desired lateral acceleration command, which can be directly generated through roll command changes, initiating a coordinated turn maneuver. The deactivation of the $\dot{y}_0$ feedback line can be accomplished through a gain scheduling process that gradually reduces the $K{\dot{y}_0}$ gain to zero as airspeed increases. The definition of the $K{\dot{y}_0}$ gain is provided in Table \ref{tab:gains}.

We implemented a similar gain scheduling strategy to reduce the angular acceleration gains with the airspeed, in response to experimental tests. These tests revealed an undesired oscillatory behavior around the pitch axis when the same hovering gains for the attitude control were used in the fast forward flight condition. This behavior can be attributed to the distinct actuators employed by the vehicle to achieve angular accelerations at various airspeeds.

For example, during hovering, the vehicle primarily relies on motor differential thrust to effect pitch and roll changes. However, at high airspeeds, pitch control is mainly achieved through elevation tilting differential, while roll control relies on aileron deflection. Furthermore, as airspeed increases, aerodynamic effects play a role in dampening the vehicle's attitude dynamics. 

For detailed numerical values of all the gains and saturation blocks associated with the error controller, specific to the dual-axis tilt rotor quad-plane, please refer to Table \ref{tab:gains} and Table \ref{tab:Saturation_values}.

\begin{table}
\centering
\renewcommand{\arraystretch}{1.4}
\begin{tabular}{C{3.5cm} C{3.5cm}} 
\hline
\multicolumn{2}{c}{\textbf{Error controller block gains}} \\
\hline
\hline
$ K_\phi$ = $K_v$ \; $s^{-1}$ & $ K_\theta$ = $K_v$ \; $s^{-1}$ \\
$ K_p$ = $4 K_v$ \; $s^{-1}$ & $ K_q$ = $4 K_v$ \; $s^{-1}$ \\
$ K_r$ = $5 K_v$ \; $s^{-1}$  & $ K_{\dot{x}}$  = 1 \; $s^{-1}$  \\
$ K_{\dot{y}}$ = 1 \; $s^{-1}$  & $ K_{\dot{z}}$  = 3 \; $s^{-1}$  \\
$K_{\dot{y}_0}$ = 1 - $K_{\dot{\Psi}_{air}}$ & $K_v = (1 - 0.03\cdot V_{a} )$ \\
\\[-1em]
\hline
\multicolumn{2}{c}{\textbf{Coordinated turn block gains}} \\
\hline
\hline
\multicolumn{2}{c}{$ K_\beta$ = 0.15 $\frac{s}{m}$} \\
\\[-1.3em]
\hline
\end{tabular}
\vspace{.1cm}
\caption{Gains used by the dual-axis tilting rotor quad-plane for the Error Controller block and the coordinated turn block. The speed $V_{a}$ is expressed in m/s and $K_{\dot{\Psi}_{air}}$ is defined in Equation \ref{coordinated_turn_block_law}.}
\label{tab:gains}
\end{table}

\begin{table}
\centering
\renewcommand{\arraystretch}{1.2}
\begin{tabular}{C{3cm}  C{1.6cm}  C{1.6cm} } 
\hline
\textbf{Saturation block} & \textbf{Min value} & \textbf{Max value} \\
\hline
$S_{\dot{x}}$ & -4 $m/s$ & 15 $m/s$  \\
\hline
$S_{\dot{y}}$ & -8 $m/s$ &  8 $m/s$  \\
\hline
$S_{\dot{z}}$ & -6 $m/s$ &  6 $m/s$ \\
\hline
$S_{\ddot{x}}$ & -3 $m/s^2$ &  3 $m/s^2$  \\
\hline
$S_{\ddot{y}}$ & -4 $m/s^2$ & 4 $m/s^2$  \\
\hline
$S_{\ddot{z}}$ & -5 $m/s^2$ & 5 $m/s^2$  \\
\hline
\end{tabular}
\vspace{.1cm}
\caption{Maximum and minimum values of the Error Controller saturation blocks depicted in Figure \ref{EC_nonlinear_unified_controller}.}
\label{tab:Saturation_values}
\end{table}

\subsection{Unified Nonlinear Control Allocation block}
The Unified Nonlinear Control Allocation block implements and solve the Control Allocation problem presented in Equation \ref{Normalized_control_problem}, using the control input array defined in Equation \ref{control_input_vector}. This block provides outputs for the commands of the physical actuators $\boldsymbol{u}_a$ defined in Equation \ref{actuators_list}, and the virtual commands $\phi_v$ and $\theta_v$. 

The inputs to this block include the current vehicle state and control input vector $\boldsymbol{x}_0$ and $\boldsymbol{u}_0$, which are required for evaluating the simplified system dynamics $\boldsymbol{f}_s(\boldsymbol{x}_0,\boldsymbol{u}_0)$. Additionally, it takes as inputs the desired linear and angular acceleration increments and the desired pitch and roll angles, $\theta_d$ and $\phi_d$. These desired pitch and roll angles are incorporated into the desired control input array $\boldsymbol{u}_d$, which is used in the control allocation problem. In our specific problem, the desired normalized control input array takes the following form:
\begin{equation}
\begin{aligned}
    \boldsymbol{u}_d = &\left( 150, \; 150, \; 150, \; 150, \right. \\
    &\;\;\; 0, \; 0, \; 0, \; 0, \\
    &\;\;\; 0, \; 0, \; 0, \; 0, \\
    &\left. \;\; 0, \; \theta_d, \; \phi_d \right) .
\end{aligned}
   \label{eq:desired_control_input} 
\end{equation}
We choose a non-zero value for the desired motor rotational speed to ensure that the motor does not come to a complete stop during flight.

\subsection{Signal Filtering}
In the Nonlinear Unified Controller scheme depicted in Figure \ref{Nonlinear_unified_controller_scheme}, we've integrated two second order Butterworth filters, labeled as $H(s$), into the feedback lines. The filter on the $\boldsymbol{\dot{y}}_0$ feedback line serve the crucial role of low-pass filtering to mitigate sensor noise that could otherwise impact the accuracy of the Control Allocation solution. Given the incremental nature of the controller, the same Butterworth filter is also applied to the current actuator state vector $\boldsymbol{x}_0$ and the current vehicle state vector $\boldsymbol{u}_0$, as described in \cite{10.2514/1.G004520}. For our specific vehicle, we utilized a second-order Butterworth filter with a cutoff frequency of $\omega_n = 13$ rad/s.

\section{Controller setup and parameters} 
\label{sec:Controller_setup_and_parameters}
\subsection{Actuators dynamics identification}
The Control Allocation problem presented in Equation \ref{Normalized_control_problem} necessitates an estimation of the current control input state for the determination of the control input solution. A model of the actuators is used to estimate the current actuator state based on the actuator commands. Compared to our prior work \cite{10.1109/ICUAS54217.2022.9836063}, we made changes in this work, such as adopting a different propeller and battery to better support high-speed forward flight. Consequently, a new system identification process, akin to the one conducted in \cite{10.1109/ICUAS54217.2022.9836063}, was executed to identify the actuator dynamics of the vehicle. The results of this system identification process are presented in Table \ref{tab:Actuators_dynamics}. 
Regarding the estimation of the last two elements of the control input array, $\theta$ and $\phi$, they are determined using a sensor based AHRS filter that runs on the main vehicle autopilot.
\begin{table}
\centering
\begin{tabular}{>{\centering}m{1.3cm}  >{\centering}m{1.5cm}  >{\centering}m{1.1cm}  >{\centering}m{.6cm} C{1cm} } 
\hline
\textbf{Actuator} & \textbf{Corner frequency $\boldsymbol{\omega_c}$} & \textbf{Damping ratio $\boldsymbol{\zeta}$} & \textbf{Rate limit} & \textbf{Delay} \\
\hline
Motor & 25 rad/s & - & - & 1 mS\\
\hline
Ailerons & 20 rad/s & - & - & 15 mS\\
\hline
Tilt elevation & 60 rad/s & 1.5 & 11.34 rad/s & 15 mS\\
\hline
Tilt azimuth & 45 rad/s & 1.6 & 9.95 rad/s & 15 mS\\
\hline
\end{tabular}
\vspace{.1cm}
\caption{Continuous time characteristics of the dual-axis tilt rotor quad-plane. The ailerons and motors dynamics are identified using a first order transfer function while a second order transfer function with rate limiter is used for the tilting dynamics.}
\label{tab:Actuators_dynamics}
\end{table}

\subsection{Thrust and torque coefficients identification}
Another crucial aspect of thrust modeling involves identifying the propeller thrust and torque coefficients, denoted as $K_p^T$ and $K_p^M$ respectively. To streamline the derivation of the system dynamics in the Appendix, we made the assumption that the inflow angle's influence on thrust and torque generation is negligible. However, even under this assumption, the propeller thrust and torque coefficients cannot be considered constant across the entire flight envelope. These coefficients still exhibit a dependence on airspeed. To capture this characteristic, a series of wind tunnel motor-propeller tests were conducted to determine the propeller thrust and torque coefficients at various airspeed. These tests were performed with the propeller disk oriented perpendicular to the airspeed direction, and the resulting coefficients are presented in Equation \ref{trust_torque_coeff}. It is worth mentioning that this model was developed based on a maximum airspeed of 20 m/s, and as such, it is not applicable to airspeed exceeding this value. 
\begin{equation}
    \begin{aligned}
        &K_p^T = 0.55e^{-5}(1 - V_{a}*0.025) \\
        &K_p^M = 0.94e^{-7}(1 - V_{a}*0.025),
    \end{aligned}
    \label{trust_torque_coeff}
\end{equation}

\begin{table}
\centering
\begin{tabular}{>{\centering}m{2.5cm}  >{\centering}m{1.6cm}  c } 
\hline
\textbf{Actuator} & \textbf{Min value} & \textbf{Max value} \\
\hline
Motors ($\Omega_i$) & 150 rad/s & 1400 rad/s  \\
\hline
Tilt elevation ($b_i$) & - 120 deg & 25 deg  \\
\hline
Tilt azimuth ($g_i$) & - 45 deg & 45 deg  \\
\hline
Ailerons ($\delta_a$) & - 25 deg & 25 deg \\
\hline
\end{tabular}
\vspace{.1cm}
\caption{Maximum and minimum physical actuator limits. For the definition of the tilting angles, please refer to Figure \ref{Reference_frames_overview}.}
\label{tab:Actuators_max_min}
\end{table}

\subsection{Control input constraint definition}
To address the optimization problem outlined in Equation \ref{Normalized_control_problem}, it is necessary to define the constraints for the normalized control input vector $\boldsymbol{u}^*$. The control input box constraints, denoted as $\boldsymbol{u}_{max}^*$ and $\boldsymbol{u}_{min}^*$, are selected as follows:
\begin{equation}
    \begin{aligned}
        & \boldsymbol{u}_{max}^* = \boldsymbol{u}_{max}/ \boldsymbol{G} \\
        & \boldsymbol{u}_{min}^* = \boldsymbol{u}_{min}/ \boldsymbol{G} \\
    \end{aligned}
\end{equation}
with
\begin{equation}
    \begin{aligned}
        & \boldsymbol{u}_{max} = \left( \Omega_1^{max}, \Omega_2^{max},  \Omega_3^{max},  \Omega_4^{max} \right.\\
        & \hspace{1.6cm} b_1^{max}, b_2^{max}, b_3^{max}, b_4^{max}\\
        & \hspace{1.6cm} g_1^{max}, g_2^{max}, g_3^{max}, g_4^{max}\\
        & \left.\hspace{1.7cm} \delta_a^{max}, \theta^{max}, \phi^{max} \right);\\
        & \boldsymbol{u}_{min} = \left( \Omega_1^{min}, \Omega_2^{min},  \Omega_3^{min},  \Omega_4^{min} \right.\\
        & \hspace{1.6cm} b_1^{min}, b_2^{min}, b_3^{min}, b_4^{min}\\
        & \hspace{1.6cm} g_1^{min}, g_2^{min}, g_3^{min}, g_4^{min}\\
        & \left.\hspace{1.7cm} \delta_a^{min}, \theta^{min}, \phi^{min} \right);\\
    \end{aligned}
    \label{eq:box_constr_definition}
\end{equation}
Where the scaling array $\boldsymbol{G}$ is defined in Equation \ref{eq:scaling_array_G} and the maximum and minimum values for the physical actuators $\Omega, b, g$ and $\delta_a$ are defined in Table \ref{tab:Actuators_max_min}. 

As for virtual commands $\theta$ and $\phi$ maximum and minimum values, these constraints take on the interpretation of aircraft attitude limits. While a constant constraint of $\phi^{max} = 40 \;deg$ and $\phi^{min} = -40 \;deg$ is set for the roll angle, the interpretation of the pitch angle constraint becomes particularly interesting. This is because an \ac{AoA} protection logic can be implemented for the calculation of the pitch angle constraint at every iteration. Consequently, the maximum and minimum values of the pitch angle are dynamically assessed at every time step using the \ac{AoA} protection algorithm, which is elaborated upon in the subsequent subsection.
This protection ensures the wing can generate maximum lift in the low to mid airspeed regime, preventing wing stall and the associated drag penalty.
This enables the effective utilization of all thrust power for vehicle forward acceleration, without diverting rotor power to contribute to vertical force generation.

\subsubsection{AoA estimation and protection}
\hfill \\
To enhance the transition control from low-speed to high-speed flight, the Unified Nonlinear Control Allocation structure can incorporate an \ac{AoA} protection algorithm.

For small roll angles and in the absence of external wind, a direct relationship can be established between the \ac{AoA} and the pitch angle, as expressed by the following equation:
\begin{equation}
  \alpha = \theta - \gamma. 
  \label{aoa_estimation}
\end{equation}
Here, $\alpha$ denotes the \ac{AoA}, and $\gamma$ represents the flight path angle, defined as follows: 
\begin{equation}
    \gamma = sin^{-1}(-\frac{\dot{z}}{V}).
    \label{gamma_definition}
\end{equation}
In Equation \ref{gamma_definition}, $\dot{z}$ signifies the vertical speed in the control reference frame, while $V$ denotes the total vehicle speed, calculated as $V = \sqrt{\dot{x}^2 + \dot{y}^2 + \dot{z}^2}$. It's important to note the negative sign in Equation \ref{gamma_definition}, which is necessary because the control reference frame's z-axis points downward.

Additionally, it's worth highlighting that the relationship between the \ac{AoA} and the pitch angle, as described in Equation \ref{aoa_estimation}, and the definition of the flight path angle $\gamma$ in Equation \ref{gamma_definition}, are also utilized in the Unified Nonlinear Control Allocation block for estimating the relation between the pitch angle and the aerodynamic forces generated by the vehicle.

Based on the relationship described in Equation \ref{aoa_estimation}, when the flight path angle $\gamma$ is known, it becomes feasible to determine a pitch angle corresponding to a specific \ac{AoA} value. Consequently, the pitch angle box constraint applied at each iteration in Equation \ref{eq:box_constr_definition} can be computed based on a desired \ac{AoA} constraint as follows:
\begin{equation}
    \begin{aligned}
    &\theta^{min} = 
    \begin{cases}
     \theta_{min}^{hard}  & \hspace{.2cm} V_{a} \leq V_{\alpha_{prot}}^{min}\\ 
     \noalign{\vskip3pt}
     max(\theta_{min}^{hard}  ,  \alpha_{min} + \gamma_0) & \hspace{.2cm} V_{a} > V_{\alpha_{prot}}^{min}
    \end{cases} \\
    &\theta^{max} = 
    \begin{cases}
     \theta_{max}^{hard}  & \hspace{.2cm} V_{a} \leq V_{\alpha_{prot}}^{min}\\ 
     \noalign{\vskip3pt}
     min(\theta_{max}^{hard}  ,  \alpha_{max} + \gamma_0) & \hspace{.2cm} V_{a} > V_{\alpha_{prot}}^{min}
    \end{cases}.
    \end{aligned}
    \label{AoA_prot_equations}
\end{equation}
Here, $V_{\alpha_{prot}}^{min}$ represents the speed threshold above which the \ac{AoA} protection algorithm becomes active. Additionally, $\alpha_{min}$ and $\alpha_{max}$ denote the desired minimum and maximum angle of attack limits. $\gamma_0$ represents the current flight path angle, while $\theta_{min}^{hard}$ and $\theta_{max}^{hard}$ are the pitch angle minimum and maximum hard limits applied across the entire flight envelope. The specific values chosen for $V_{\alpha_{prot}}^{min}$, $\theta_{min}^{hard}$, $\theta_{max}^{hard}$, $\alpha_{min}$, and $\alpha_{max}$ for the dual-axis tilting rotor quad-plane are provided in Table \ref{tab:aoa_constraint_values}.

\begin{table}
\centering
\renewcommand{\arraystretch}{1.2}
\begin{tabular}{>{\centering}m{2.5cm}  c } 
\hline
$ V_{\alpha_{prot}}^{min}$ & 6 \; m/s  \\
\hline
$\theta_{min}^{hard}$ & -20 \; deg  \\
\hline
$\theta_{max}^{hard}$ & 80 \; deg  \\
\hline
$\alpha_{min}$ & - 5 \; deg  \\
\hline
$\alpha_{max}$ & 15 \; deg  \\
\hline
\end{tabular}
\vspace{.1cm}
\caption{Dual-axis tilting rotor quad-plane parameters required for the evaluation of the pitch angle constraint in Equation \ref{AoA_prot_equations}}
\label{tab:aoa_constraint_values}
\end{table}

\subsection{Resolution of the control optimization problem}

For the solution of the normalized control problem presented in Equation \ref{Normalized_control_problem}, we utilized the \ac{SQP} algorithm. This gradient-based iterative optimization method demonstrated the shortest runtime and the fastest solution convergence for our specific problem among various available optimization algorithms.

A detailed derivation of the \ac{SQP} algorithm is beyond the scope of this paper. Interested readers can refer to the work of K. Schittkowski, where the \ac{SQP} algorithm is thoroughly explained \cite{NLQPL}.

One of the primary advantages of using the SQP algorithm is its compatibility with the MATLAB \textit{fmincon} function \footnote{\href{https://nl.mathworks.com/help/optim/ug/constrained-nonlinear-optimization-algorithms.html}{\color{blue}https://nl.mathworks.com/help/optim/ug/constrained-nonlinear-optimization-algorithms.html}}. Furthermore, the \textit{fmincon} function seamlessly integrates with the MATLAB Coder toolbox\footnote{\href{https://nl.mathworks.com/products/matlab-coder.html}{\color{blue}https://nl.mathworks.com/products/matlab-coder.html}}. This compatibility facilitates the smooth transition of the same \ac{SQP} algorithm developed in Simulink for simulation to practical testing on a real flying drone. This transition can be achieved by employing an external companion computer, such as a Raspberry Pi 4 running a Linux OS. The C code generated by MATLAB can be executed onboard the Raspberry Pi 4. The companion computer receives the desired acceleration increment commands and vehicle states from the primary flight computer through UART communication. Once the code returns the actuator solution, these commands are forwarded to the primary flight computer using the same UART communication channel.

To ensure that the Unified Nonlinear Control Allocation algorithm runs in real-time at a high refresh rate, we made a modification to the generated C code. Initially, the MATLAB generation toolbox allowed only a maximum number of iterations or function evaluations to be set. We enhanced the generated C code by introducing a maximum computational time constraint for the actuator solution, limited to 5 milliseconds. This modification ensures that when the \ac{SQP} algorithm reaches a runtime of 5 milliseconds, the computed solution from the last available iterative step is selected. While this constraint may result in a less accurate solution, it guarantees that we achieve an actuator solution with a frequency of at least 200 Hz.

\subsection{Choice of the initial point for the optimizer}

To conclude the setup of the optimization problem, another crucial aspect to consider is the choice of the initial actuator solution as a starting point for the optimization process. This parameter holds significant importance because an incorrect initialization of the initial control input array can potentially lead to the computation of a sub-optimal final control input solution. This situation can occur due to the non-global convexity of the cost function or due to the premature stop of the optimization process as a consequence of an iteration limit or a maximum run-time constraint.

To determine the most suitable initial control input array $\boldsymbol{u}_{init}$ to initiate the optimization problem of Equation \ref{Normalized_control_problem}, we conducted a comprehensive statistical analysis. We carried out 500 realizations of desired accelerations and flying vehicle states, progressively increasing the maximum allowable iterations for solving the optimization problem. For each realization of flight state, desired accelerations and current control input $\boldsymbol{u}_0$, we compared the results obtained using a random valid initial actuator value $\boldsymbol{u}_{rand}$ with those obtained using the current actuator state $\boldsymbol{u}_0$ as a starting point for the optimizer. We then evaluated the solution quality based on the final cost function value and the norm of acceleration residuals.

\begin{figure*}
	\centering
	\includegraphics[scale  = .57]{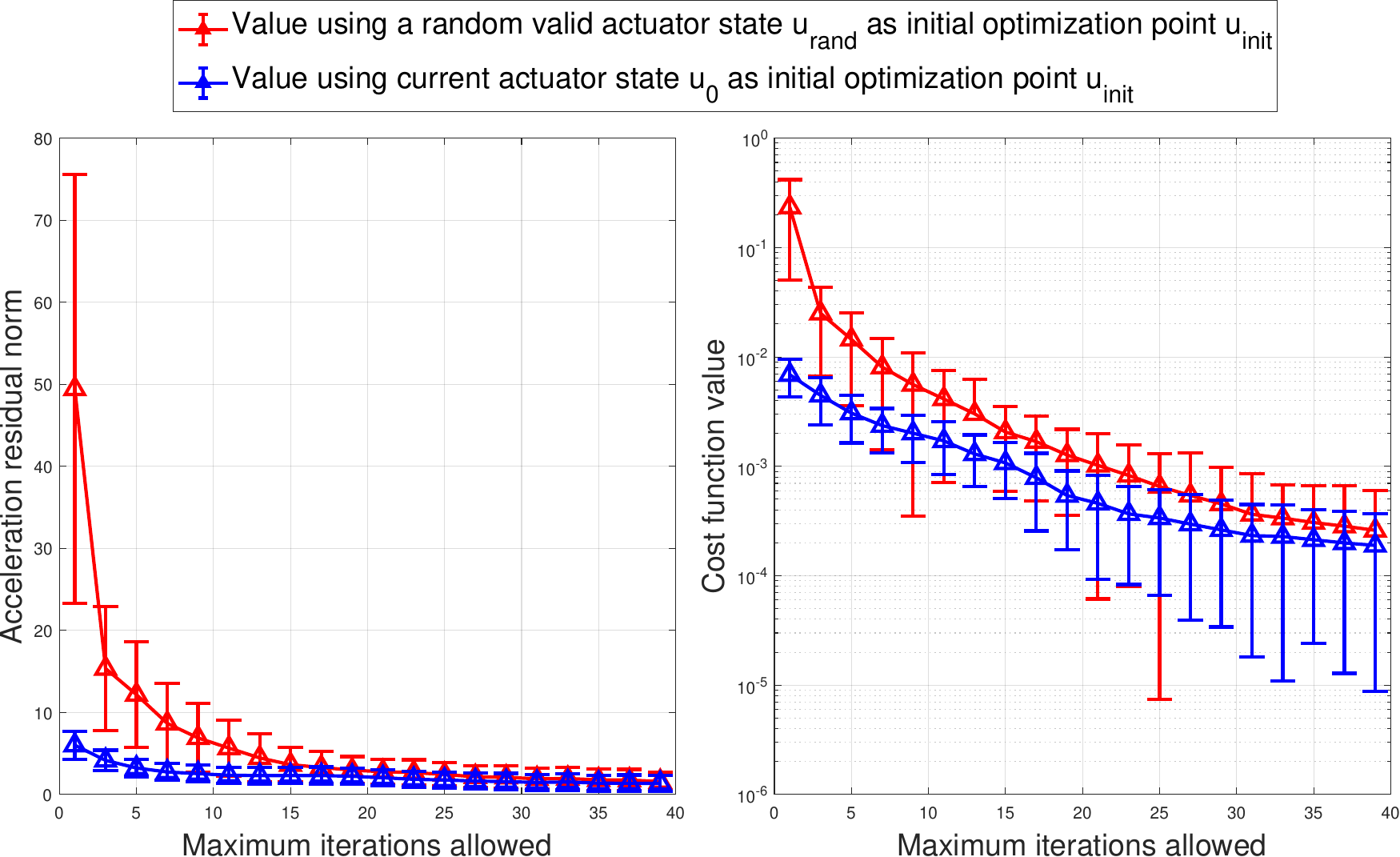}
	\caption{Norm of Acceleration Residuals and Final Cost Function for Solutions Obtained via the Nonlinear Control Allocation Problem in Equation \ref{Normalized_control_problem} with Increasing maximum Iterations. For each iteration, 500 realizations of desired accelerations and current vehicle states were evaluated, employing random and current actuator state as initial guesses for the optimizer.}
	\label{fig:stat_analysis_increasing_number_of_iter}
\end{figure*}

The results, as shown in Figure \ref{fig:stat_analysis_increasing_number_of_iter}, unequivocally illustrate that commencing the optimization process from the current actuator state results in rapid convergence compared to starting from a random initial point. Consequently, we have chosen to employ the current actuator state $\boldsymbol{u}_0$ as the initial actuator state for the optimization process $\boldsymbol{u}_{init}$.

\section{Flight test experiment and results}
\label{sec:Flight_test_experiment_and_results}
In this section, we will present and discuss flight test data obtained from two flights of the dual-axis tilting rotor quad-plane shown in Figure \ref{Prototype_figure}, operating with the previously introduced Unified Incremental Nonlinear Controller.
\subsection{Experimental setup}
The dual-axis tilting rotor vehicle used in the experiments has a takeoff mass of 2.5 kg and possesses the physical characteristics outlined in Table \ref{tab:physical_properties_vehicle}.

\begin{figure*}
	\centering
	\includegraphics[scale  = .5]{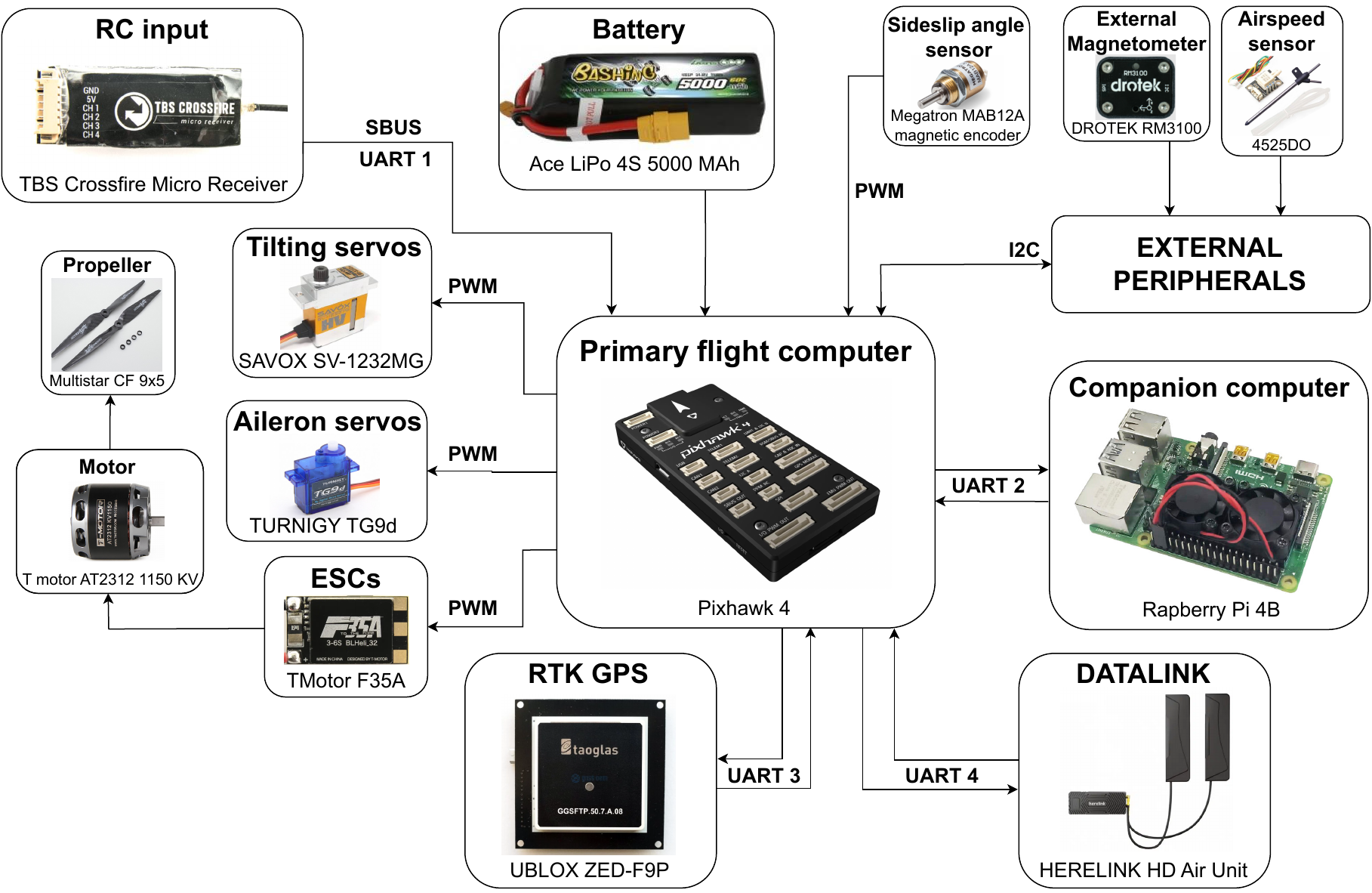}
	\caption{Scheme of the dual-axis tilting rotor quad-plane avionics.}
	\label{fig:hardware_scheme}
\end{figure*}

The avionics components of the vehicle are illustrated in Figure \ref{fig:hardware_scheme}. For state estimation, reference generation, and ground station communication, we employed the open-source Paparazzi UAV \footnote{\href{https://wiki.paparazziuav.org/}{\color{blue}https://wiki.paparazziuav.org/}} autopilot, running on the Primary Flight Computer Pixhawk 4.

Regarding the operating system on the Raspberry Pi 4, we opted for the Raspberry Pi OS, which is a Unix-like operating system based on the Debian Linux distribution.

\begin{table}
\centering
\renewcommand{\arraystretch}{1.3}
\begin{tabular}{P{3.5cm}  P{3.5cm} } 
\hline
\multicolumn{2}{c}{\bfseries Vehicle physical characteristics} \\
\hline 
$I_{xx}$ & 0.156 \; $Kg\cdot m^2$\\
\hline 
$I_{yy}$ & 0.161 \; $Kg\cdot m^2$\\
\hline
$I_{zz}$ & 0.259 \; $Kg\cdot m^2$\\
\hline
$Mass$ & 2.44 \; $Kg$\\
\hline
$S$ & 0.43 \; $m^2$\\
\hline
$\bar{c}$ & 0.3\; $m$\\
\hline
$b$ & 1.4 \;$m$\\
\hline
$l_1$ & 0.228 \;$m$\\
\hline
$l_2$ & 0.228 \;$m$\\
\hline
$l_3$ & 0.38 \;$m$\\
\hline
$l_4$ & 0.38 \;$m$\\
\hline
$l_z$ & 0 \;$m$\\
\hline
\end{tabular}
\vspace{.1cm}
\caption{Vehicle physical characteristics. The inertia elements have been accurately estimated through a meticulous CAD modeling process that encompasses each hardware component of the vehicle. For the definition of the rotor displacement $l_i$ please refer to Figure \ref{fig:Assumptions_and_notation}.}
\label{tab:physical_properties_vehicle}
\end{table}

\subsection{Flight test design}
To assess the capabilities of the presented Unified Incremental Nonlinear Controller, two distinct autonomous maneuvers were designed and executed aboard the vehicle. Manual control was utilized for the takeoff and landing phases to ensure safety, while autonomous flight test maneuvers were initiated from a hovering state at a safe altitude.

The first maneuver involved achieving specific vertical, forward, and lateral accelerations. In the second maneuver, we replicated the first one but introduced an additional challenge. Throughout the entire duration of the maneuver, we imposed a desired pitch angle of $\theta_d = 25 deg$. This specific maneuver aimed to test the controller's behavior when a desired vehicle attitude is requested during a flight phase in which the vehicle does not have full 6 \ac{DOF} control capabilities, such as during fast forward flight. Under such conditions, the controller is expected to prioritize pitch control to achieve the desired vertical acceleration, effectively disregarding the requested pitch angle.

The hovering capabilities of the vehicle were not intentionally tested, as they were extensively examined in our previous work \cite{10.1007/s10846-023-01865-8}.

\begin{figure*}
	\centering
	\includegraphics[scale  = .58]{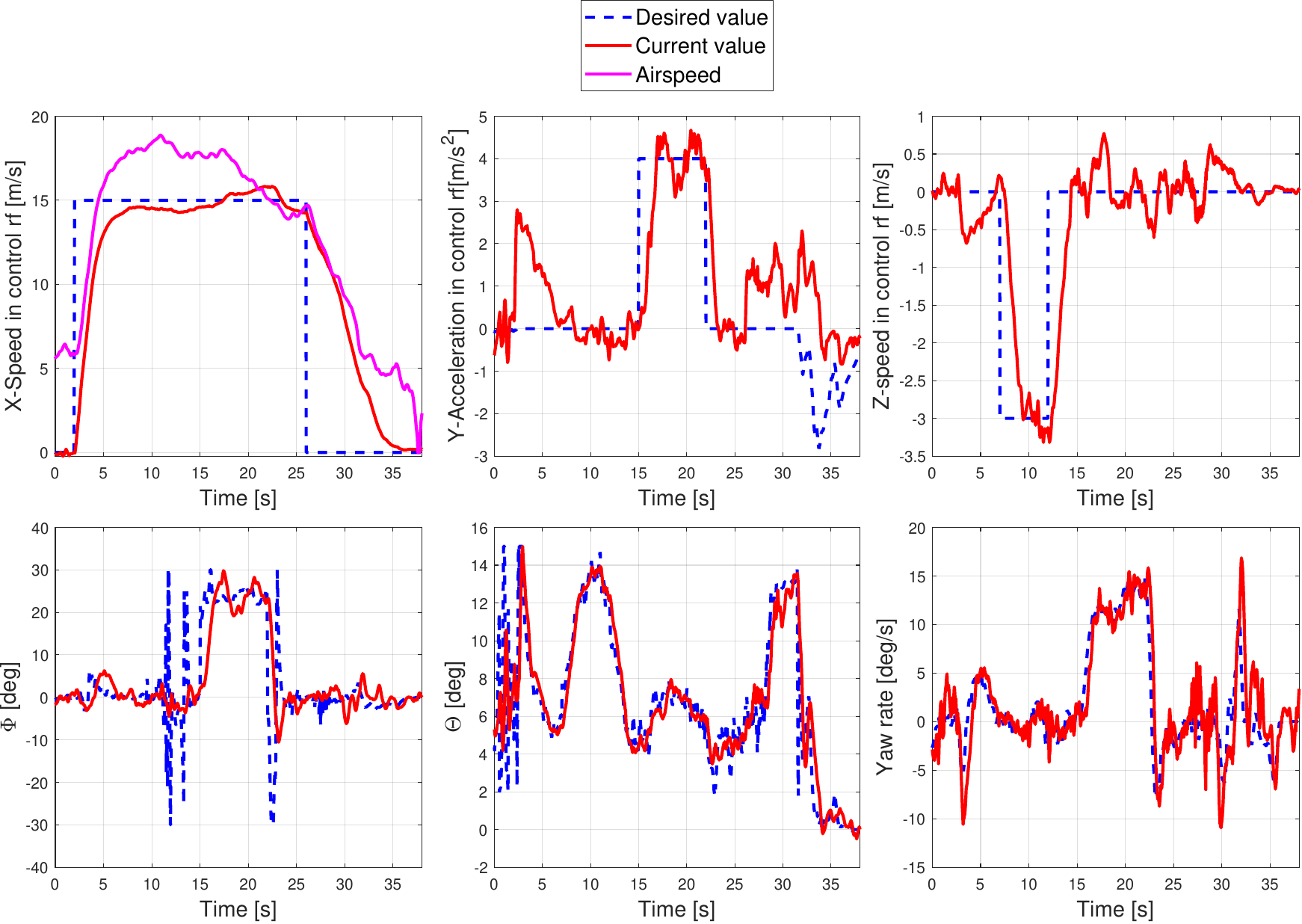}
	\caption{Evolution of vehicle states and desired state values for the first maneuver. At the top of the figure, you can see the forward speed, lateral acceleration, and vertical acceleration over time. In the lower section, the evolution of roll, pitch, and yaw rate is displayed.}
	\label{fig:first_maneuver_speed_attitude}
\end{figure*}

\begin{figure*}
	\centering
	\includegraphics[scale  = .58]{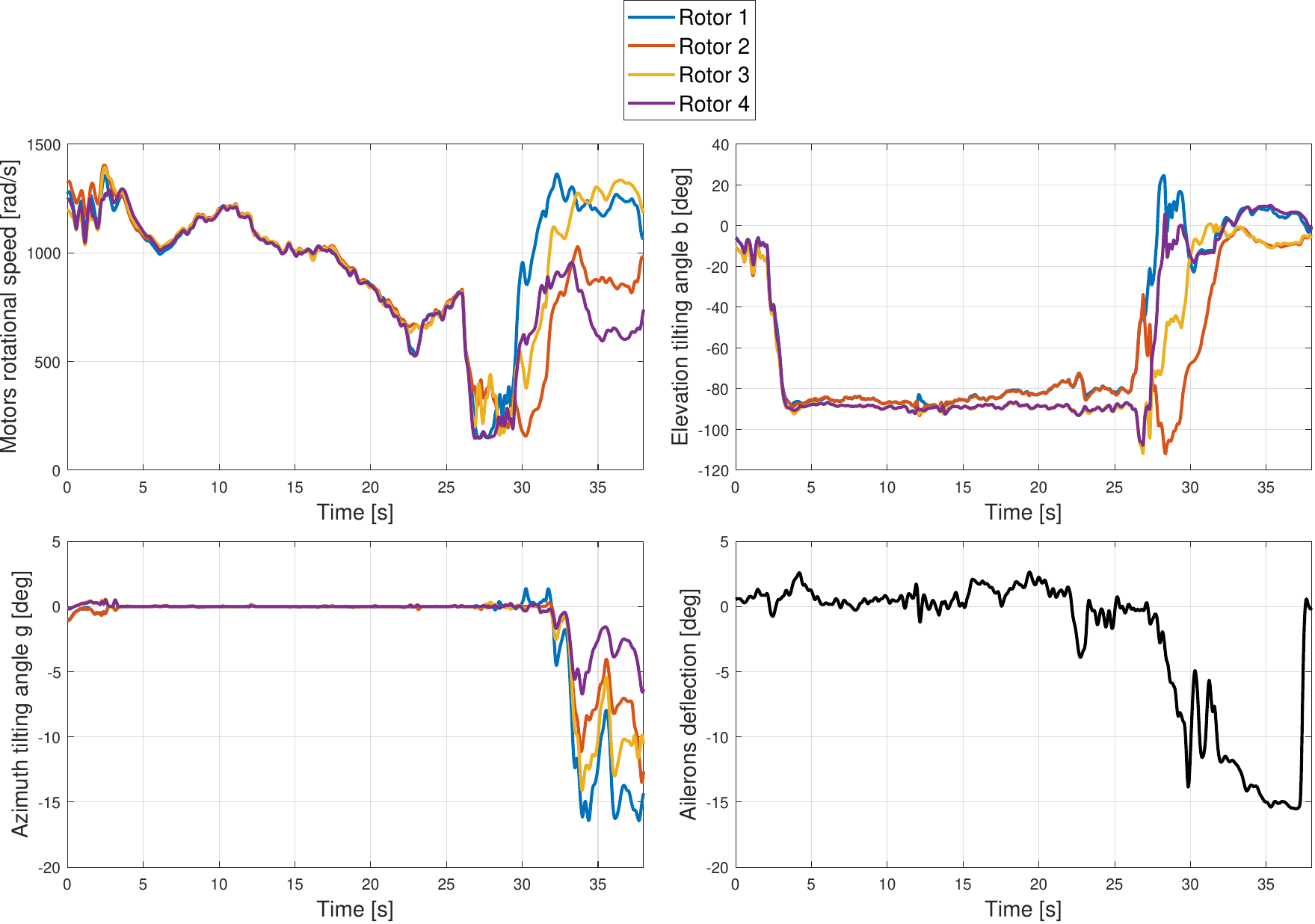}
	\caption{Actuators evolution for the first maneuver. Note that the actuators command are directly computed by the Unified Nonlinear Control Allocation block.}
	\label{fig:first_maneuver_actuator_evolution}
\end{figure*}

\begin{figure*}
	\centering
	\includegraphics[scale  = .58]{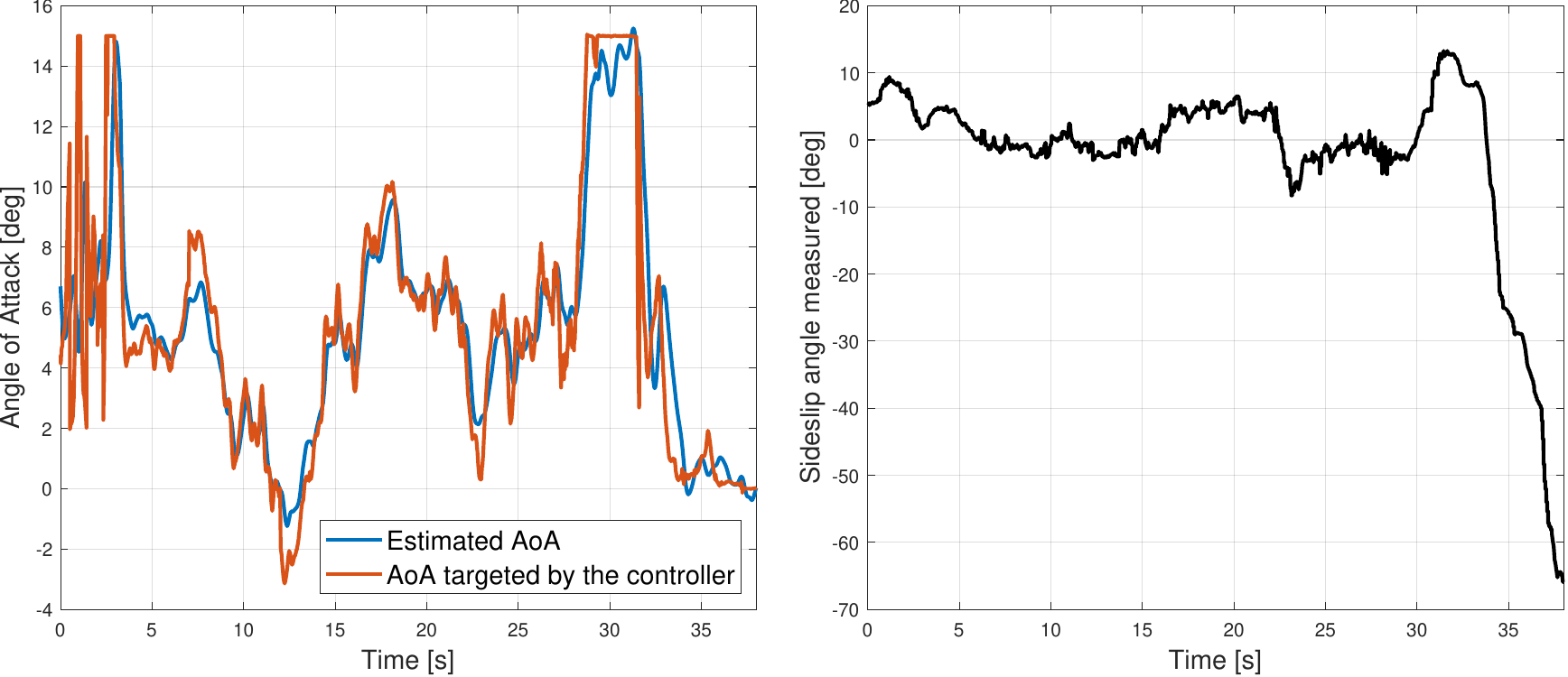}
	\caption{Evolution of the estimated \ac{AoA} and the measured sideslip angle during the first maneuver.}
	\label{fig:first_maneuver_AoA_and_sideslip}
\end{figure*}

\begin{figure*}
	\centering
	\includegraphics[scale  = .58]{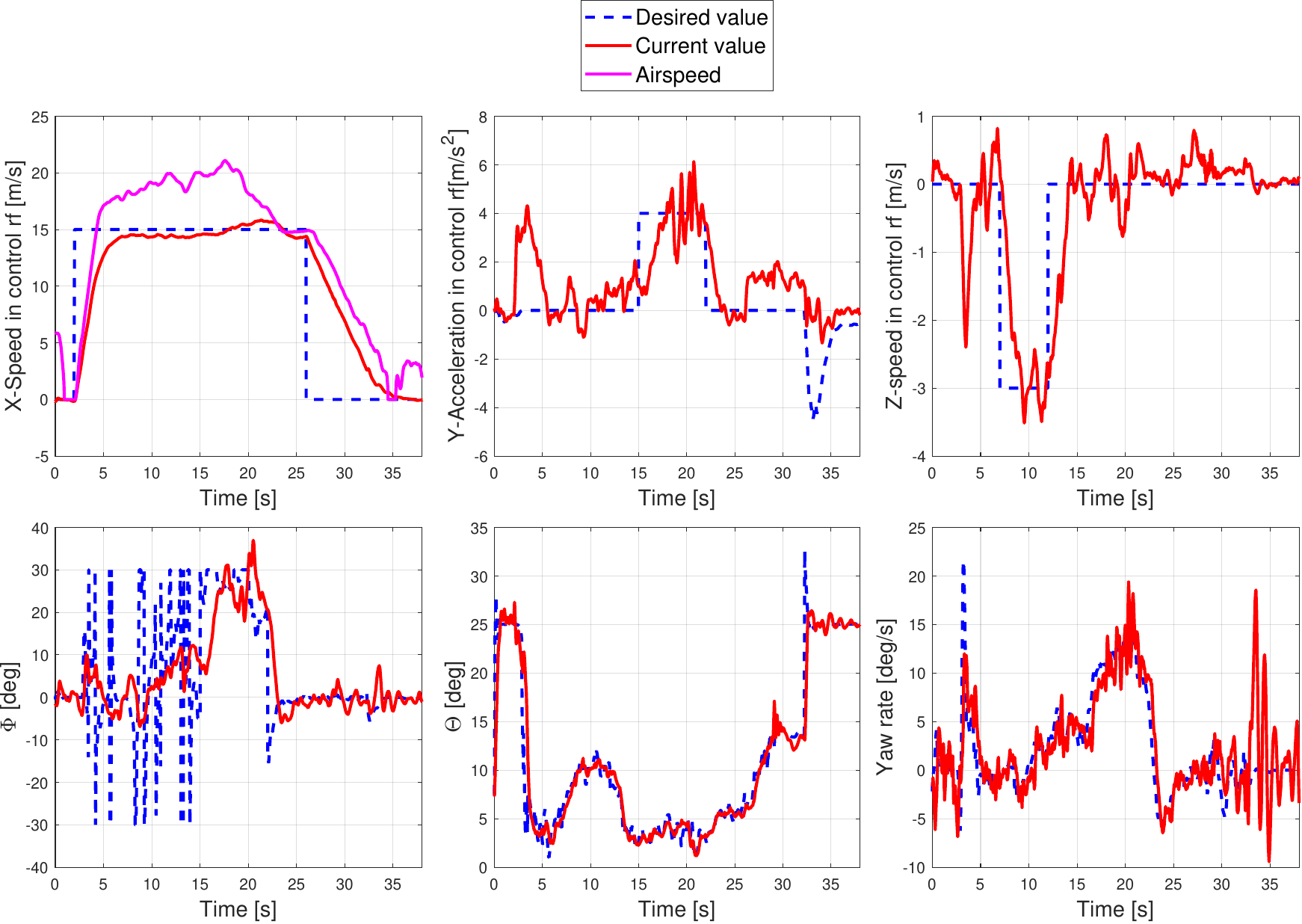}
	\caption{Evolution of vehicle states and desired state values for the second maneuver. At the top of the figure, you can see the forward speed, lateral acceleration, and vertical acceleration over time. In the lower section, the evolution of roll, pitch, and yaw rate is displayed.}
	\label{fig:second_maneuver_speed_attitude}
\end{figure*}

\begin{figure*}
	\centering
	\includegraphics[scale  = .58]{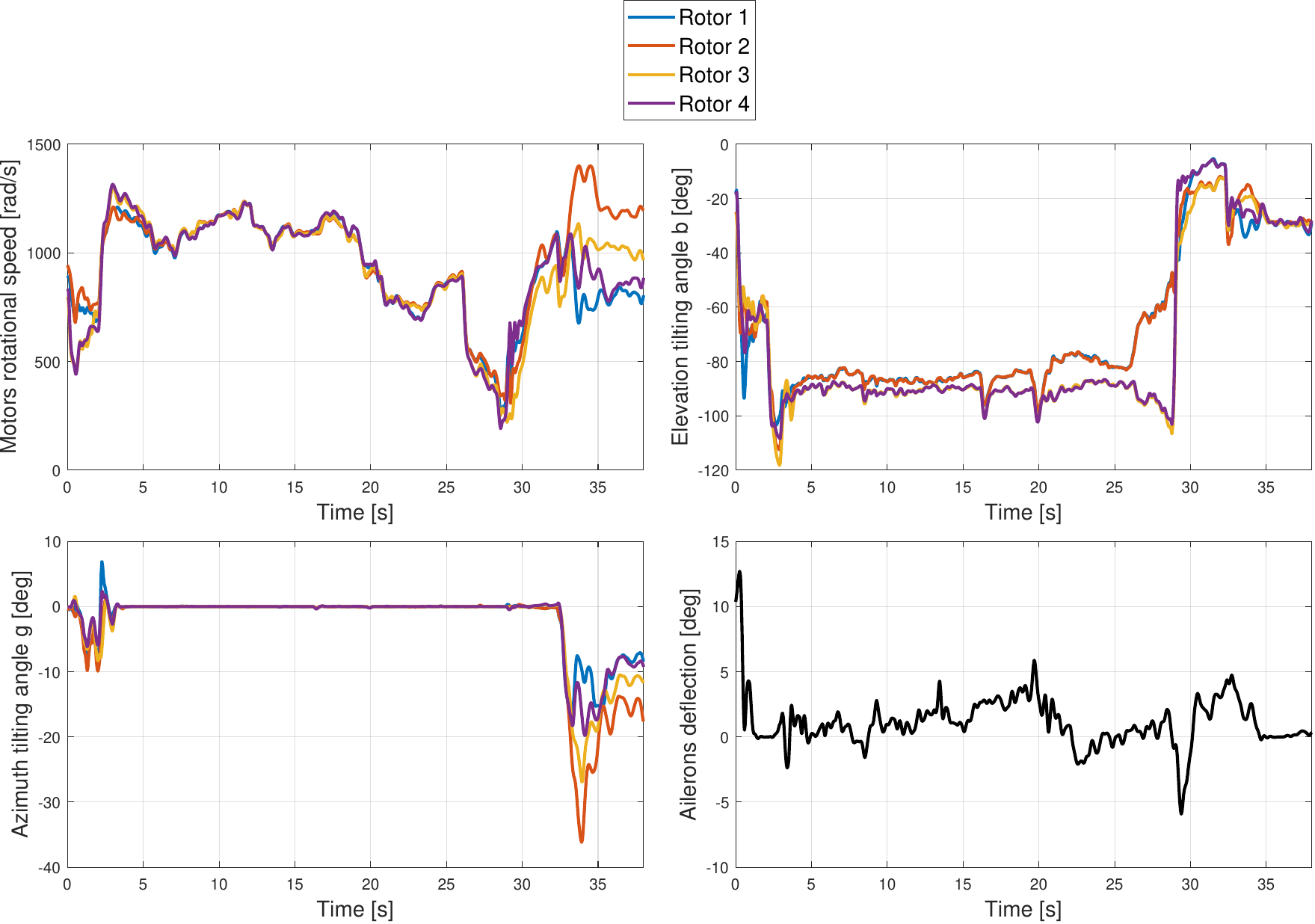}
	\caption{Actuators evolution for the second maneuver. Note that the actuators command are directly computed by the Unified Nonlinear Control Allocation block.}
	\label{fig:second_maneuver_actuator_evolution}
\end{figure*}

\begin{figure*}
	\centering
	\includegraphics[scale  = .58]{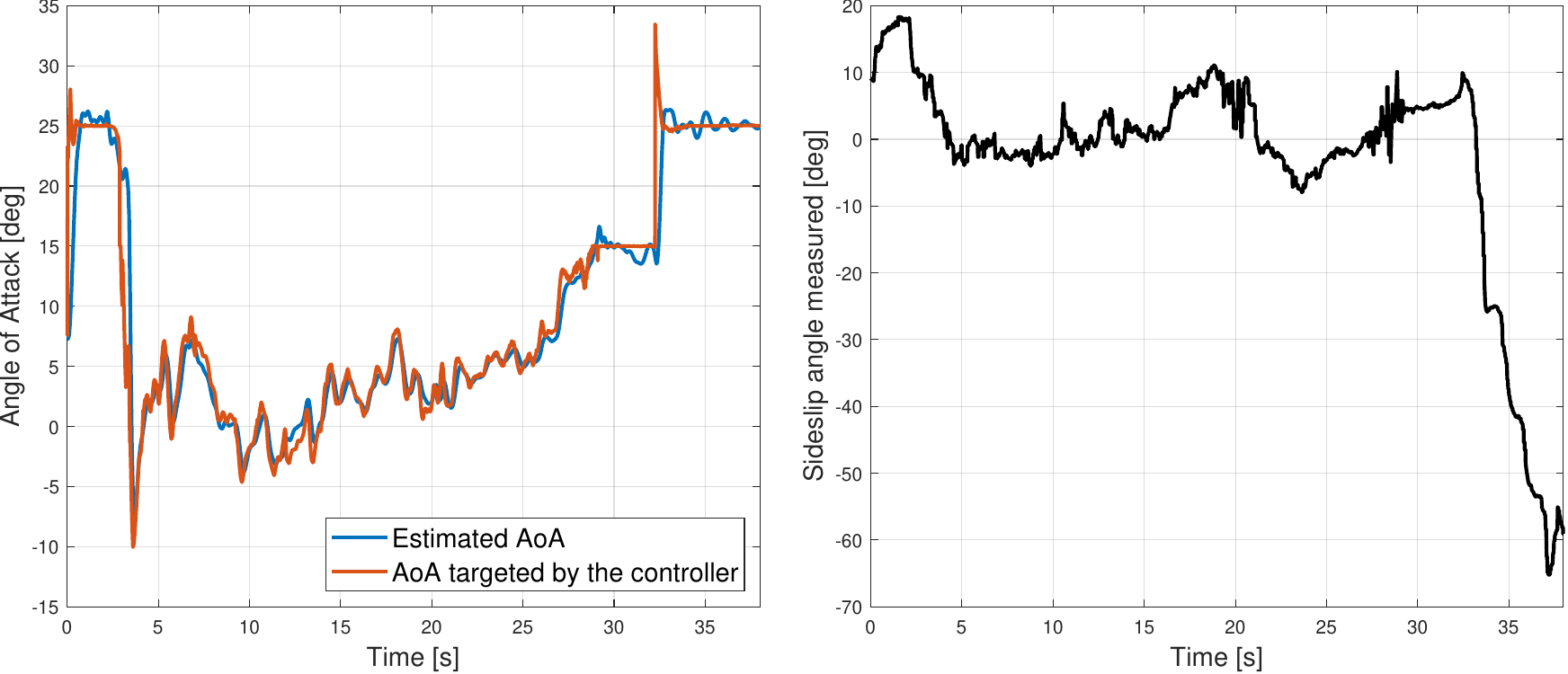}
	\caption{Evolution of the estimated \ac{AoA} and the measured sideslip angle during the second maneuver.}
	\label{fig:second_maneuver_AoA_and_sideslip}
\end{figure*}

\begin{figure*}
	\centering
	\includegraphics[scale  = .58]{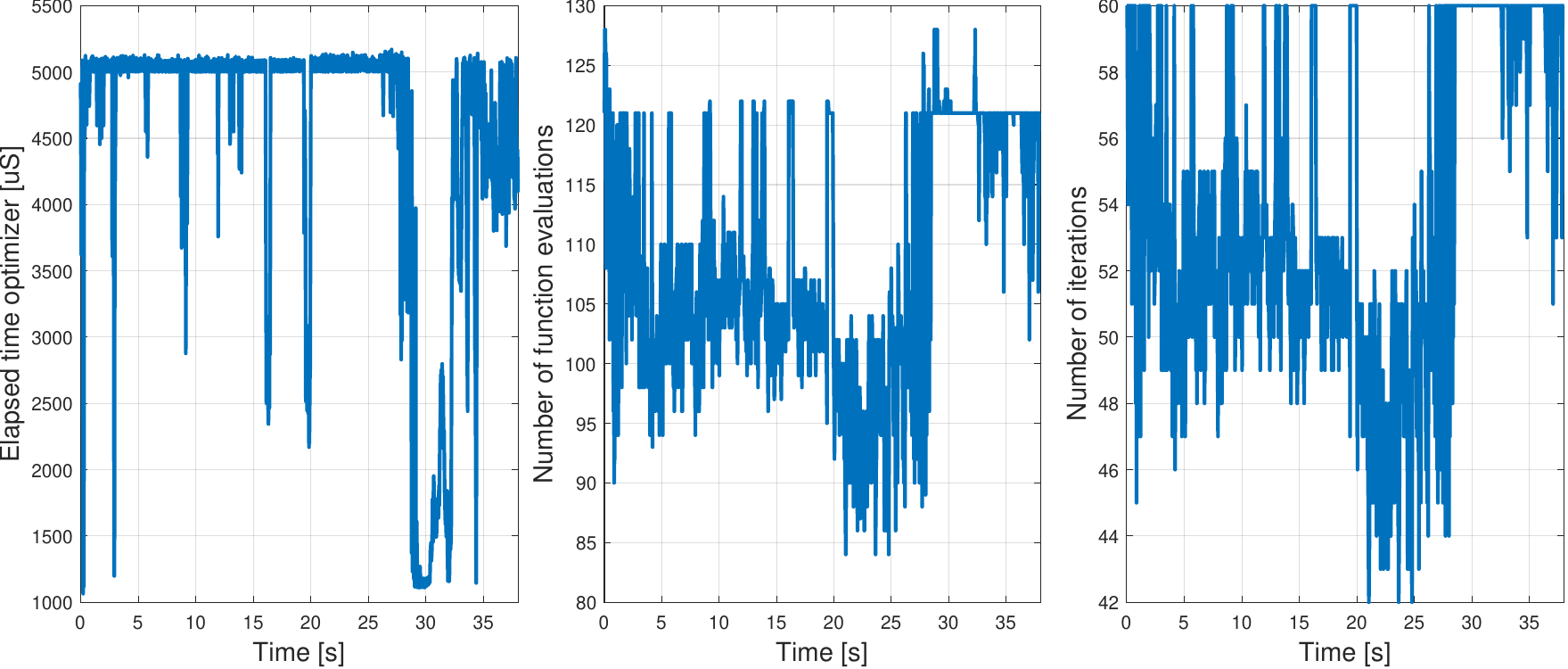}
	\caption{Run-time, function evaluations, and the number of iterations required by the Unified Nonlinear Control Allocation algorithm to compute the actuator solution during the second maneuver.}
	\label{fig:second_maneuver_optimizer_performance}
\end{figure*}

\subsection{Flight test results}

After analyzing the flight test results of the two maneuvers reported from Figure \ref{fig:first_maneuver_speed_attitude} to Figure \ref{fig:second_maneuver_optimizer_performance}, several observations can be made:
\begin{itemize}[left=0pt, itemsep=0em]
  \item As evident from Figure \ref{fig:first_maneuver_speed_attitude} and \ref{fig:second_maneuver_speed_attitude}, it is clear that all desired speeds, lateral accelerations, attitude angles, and yaw rates are well tracked. Minimal error is observed between the desired and current states throughout both the first and second maneuvers.
  
  \item The Control Allocation problem is consistently solved correctly, as evidenced by the computed actuator solution in Figure \ref{fig:first_maneuver_actuator_evolution} and Figure \ref{fig:second_maneuver_actuator_evolution}. The actuator solution remains smooth throughout the maneuvers for all actuators (motors, tilt, and ailerons) with no noticeable oscillations or abrupt changes. Furthermore, it's evident how the controller adapts to different vehicle dynamics during attitude control. At low speeds, pitch and roll actions are primarily controlled through motor thrust commands, while yaw rate is controlled through tilting action. In fast forward flight, the pitch angle is controlled through differential elevation tilt commands, and roll actions are controlled by the ailerons. As for the yaw rate, in forward flight, it is controlled with differential motor thrust commands.
  
  \item As demonstrated in Figure \ref{fig:first_maneuver_AoA_and_sideslip} and Figure \ref{fig:second_maneuver_AoA_and_sideslip}, the \ac{AoA} protection algorithm consistently maintains the \ac{AoA} within the range of $\pm$ 15 degrees when the airspeed exceeds 6 m/s.
  
  \item Thanks to the coordinated turn block, which generates the yaw rate reference as expressed in equation \ref{coordinated_turn_block_law}, the turn maneuvers are effectively executed, and the sideslip angle is consistently minimized. The sideslip angle remains within 10 degrees during the forward flight phase of both maneuvers, as evident from Figure \ref{fig:first_maneuver_AoA_and_sideslip} and Figure \ref{fig:second_maneuver_AoA_and_sideslip}.
  
  \item The transition from hovering to forward flight is effectively managed, with minimal altitude loss. During transition, the vertical speed is initially controlled by motor thrust and then, as airspeed increases, by pitch actions.
  
  \item In the second maneuver, the desired pitch angle of 25 degrees is appropriately applied during hovering and low airspeed flight phases, allowing for full 6 \ac{DOF} control. In fast forward mode, where the vehicle has only 4 \ac{DOF}, the desired pitch angle is smoothly discarded in favor of \ac{AoA} protection and lift generation control.
  
  \item The run-time for the optimizer solving the Control Allocation problem consistently remains below or equal to 5 milliseconds, as shown in Figure \ref{fig:second_maneuver_optimizer_performance}, which displays the optimizer performance during the second maneuver. The controller maintains an average update rate of 224 Hz in the first maneuver and 222 Hz in the second one, ensuring minimal delay in the control loop.
  
    \item Lateral tilting commands are not used in fast forward flight to achieve lateral acceleration, in favor of roll angle commands. Regarding roll angle command generation, a jumping and saturating solution of the roll angle is computed in the first part of the second maneuver, mainly due to the low or negative \ac{AoA} estimated. 

    As the maneuver progresses, this problem gradually diminishes as the \ac{AoA} estimation increases during the turn. This issue is particularly troublesome because when the desired \ac{AoA} is negative, the associated desired roll angle for achieving a desired lateral acceleration becomes inverted. For instance, in the case of a negative \ac{AoA}, to generate a positive lateral acceleration, the controller would request a negative roll angle ($\phi$). Such behavior is undesirable, especially considering that commands for the yaw rate are generated using a feed forward component of the roll angle and are only valid under the assumption of a positive \ac{AoA}.

    The likely root cause of this problem could be the inaccuracy of the \ac{AoA} estimation, which relies on certain assumptions that may not be entirely valid during flight. These assumptions include the absence of external wind and minimal roll angle values. To address this issue, a more accurate in-flight estimation of the \ac{AoA} would be required. Alternatively, a potential workaround for this issue could involve imposing a constraint on the vertical acceleration to ensure a consistently positive \ac{AoA} computation.
    
    Unfortunately, due to time constraints, we were unable to conduct a new flight test to implement this potential workaround. However, we plan to include this in a follow-up study.
\end{itemize}

For interested readers, a video of the experiment, which includes both the first and second maneuver, is available at the following link: \href{https://youtu.be/80Qk2V_xwmw}{\color{blue}https://youtu.be/80Qk2V\_xwmw}.

\section{Limitations of the Unified Incremental Nonlinear Controller}
\label{sec:Limitations_of_the_Unified_Incremental_Nonlinear_Controller}
Even if the presented Unified Incremental Nonlinear Controller was capable of smoothly controlling the dual-axis tilting rotor quad-plane, some limitations can be still identified. Those limitations are still currently a source of research and are the following:
\begin{itemize}[left=0pt, itemsep=0em]
    \item \textbf{Limited Actuator Prioritization:} The current cost function definition in Equation \ref{Normalized_control_problem} incorporates two weighting matrices, $\boldsymbol{W}_u$ and $\boldsymbol{W}_\nu$, designed to either penalize or prioritize specific actuator usage and desired acceleration elements. Nevertheless, the current configuration lacks the capacity to prioritize the utilization of a particular actuator for achieving a specific component of the desired acceleration. In other words, it is currently not feasible to assign, for example, exclusive priority to the back rotors elevation tilt for achieving the roll rate while primarily utilizing the front rotors elevation tilt for pitch rate control. Instead, the system allows only for prioritizing a specific set of rotors elevation tilts for the allocation of all desired accelerations. This limitation became relevant as we noted that the impact of the back tilting rotors on the vehicle's aerodynamics was relatively minor. It would have been more efficient to allocate the roll rate in forward mode by employing the back rotor's elevation tilt, while maintaining the use of the front rotor's elevation tilt for pitch rate control. However, this is not currently possible within the existing configuration. Consequently, we opted to incorporate ailerons on the vehicle to enable proper and efficient turns in high airspeed conditions.
    
    \item \textbf{Inability to Account for Different Actuator Bandwidths:} The Control Allocation algorithm in Equation \ref{Normalized_control_problem} assumes infinite bandwidth for all actuators to determine the actuator solution. However, in our system, various actuators possess different bandwidths, as evident in Table \ref{tab:Actuators_dynamics}. Moreover, in addition to the physical actuators, we also produce commands for the roll and pitch angles, which have significantly slower dynamics compared to the physical actuators. Consequently, some actuator commands reach their desired values faster than others, resulting in transient undesired acceleration components that require correction. While one approach to mitigate this issue is to use slower error controller gains to reduce actuator commands and alleviate the actuator dynamics mismatch, it is essential to explore more effective methods to account for the physical properties of each actuator.  
    
    \item \textbf{Local minimum in the cost function:} The Nonlinear Control Allocation problem presented in Equation \ref{Normalized_control_problem} utilizes a nonlinear quadratic cost function based on the vehicle dynamics. If the cost function is not globally convex across its entire domain, there is a risk that the optimization process might converge to a local cost function minimum instead of a global minimum. Unfortunately, due to the complexity of the cost function, studying its global convexity is a challenging task. During the flight test and multiple simulations, we did not encounter any significant issue associated with local cost function minima. The optimizer consistently provided solutions that achieved the desired accelerations with minimal residuals. Nonetheless, in our previous work \cite{10.1007/s10846-023-01865-8}, we identified the presence of a few local minimum points in the cost function, in the hovering scenario. Although these local minima did not impact the optimization process's performance and results, it is prudent to conduct a more detailed analysis of the cost function's shape in future research. This analysis will assist in recognizing and mitigating potential local minima, thereby enhancing the optimization process's reliability. 
\end{itemize}

\section{Conclusion}
\label{sec:Conclusions}
Even with an imperfect aerodynamic model, the ability to effectively and accurately control the vehicle has been demonstrated. This achievement can be largely attributed to the incremental nature of the control laws, which can readily compensate for any constant offsets in the aerodynamic model parameters.

The \ac{AoA} protection algorithm played a crucial role in safeguarding the vehicle's pitch angle, minimizing the risk of wing stall, especially during transition phases. The vehicle's transition capabilities, coupled with the ability to produce lift and thrust simultaneously, have led to improved performance and minimal altitude loss during these critical flight phases.

Pitch and roll angles were effectively utilized during fast forward flight to control vertical speed and lateral acceleration, respectively. Moreover, the controller adeptly discarded desired attitude commands when transitioning between hovering and high-speed configurations, where only 4 \ac{DOF} are effectively controlled. Furthermore, the yaw rate commands generated by the coordinated turn block proved effective in managing turns while minimizing sideslip.

During the flight test phase, inaccuracies in estimating the \ac{AoA} for a limited portion of the maneuver resulted in the generation of undesirable "jumping" roll angle commands. This occurred because during that specific time frame, the error controller was demanding a significant downward acceleration. Consequently, the controller aimed to achieve an \ac{AoA} that fluctuated between small positive and negative values. In future research, we plan to address this issue through one of two methods. The first approach involves improving the accuracy of \ac{AoA} estimation. Alternatively, we are considering the implementation of an active constraint algorithm that limits the desired vertical acceleration of the vehicle. This constraint would ensure that the \ac{AoA} never reaches negative values, automatically preventing the controller from generating abrupt roll angle commands to target lateral acceleration.

In future work, we also aim to further enhance vehicle performance through accurate system identification models that consider propeller-inflow angle and the interactions between the wing and rotor-induced inflow. This will possibly enable the vehicle to achieve roll rate control in fast forward flight solely through tilting commands, potentially eliminating the need for ailerons. 

Additionally, we are currently in the process of investigating fault-tolerant control strategies for the overactuated vehicle. Our research involves developing logic that can effectively manage the loss of one or multiple actuators within the Unified Nonlinear Control Allocation block.

Further research efforts will also focus on refining the cost function to reduce the presence of local minima and incorporating mechanisms to penalize or reward the use of specific actuators to achieve certain elements of the desired accelerations vector.

\bibliography{bibliography_file}
\bibliographystyle{plainurl}



\appendix

\section{Mathematical model of the dual-axis tilting rotor quad-plane}
In this section, we will derive the mathematical model of the dual-axis tilting rotor quad-plane depicted in Figure \ref{Prototype_figure}. In our previous work \cite{10.1007/s10846-023-01865-8}, we have already derived the equation of motion for the vehicle. However, for this work, we needed to make a few modifications to that mathematical model.

In this study, the equations of motion are expressed in a new reference frame called the control reference frame. The derivation of the equation of motion in the control reference frame was necessary to simplify the model used by the Unified Nonlinear Control Allocation algorithm for determining the actuator solution. Additionally, we introduced ailerons for the vehicle, and therefore, we had to model them as well.

\subsection{Reference frames and notation} 
\begin{figure}
	\centering
	\includegraphics[scale  = .19]{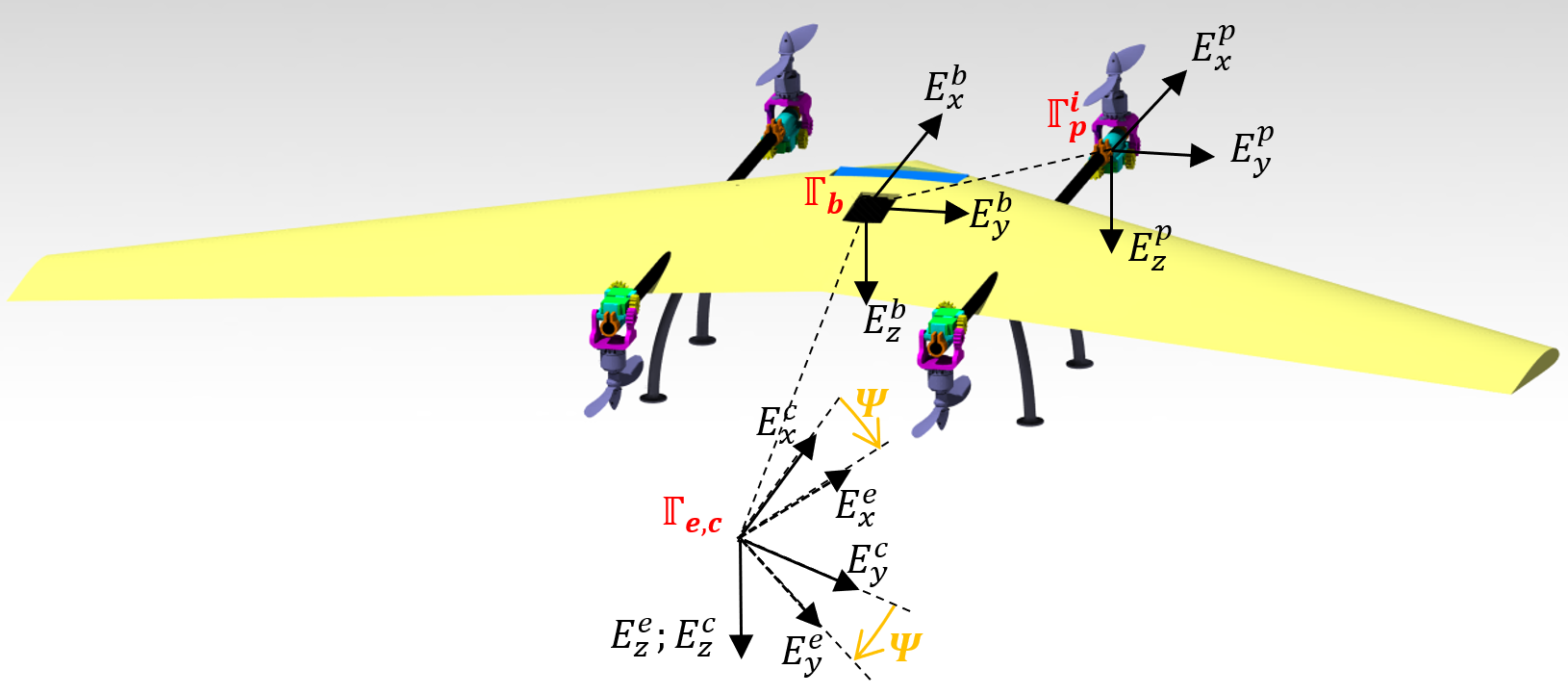}
	\caption{Overview of the Earth, Control, Body and Propeller reference frames.}
	\label{Reference_frames_overview}
\end{figure}

After establishing the rotor numbering and motor spinning direction in Figure \ref{fig:Assumptions_and_notation}, we can now introduce the various reference frames utilized for characterizing the vehicle dynamics:
\begin{itemize}

    \item Earth Frame $\Gamma_e$ : Origin on the Earth surface, $x_e$ points north, $y_e$ points east and $z_e$ axis points towards the center of the Earth. 

    \item Control Frame $\Gamma_c$ : Origin in the airplane COG, $x_c$ contained in the intersection between the symmetry plane and the $x_e$, $y_e$ plane, $z_c$ axis pointing towards the center of Earth and $y_c$ pointing in such a way to complete a right-handed orthonormal frame. 
        
    \item Body Frame $\Gamma_b$ : Origin in the airplane COG, $x_b$ axis in the vehicle plane of symmetry and pointing to the nose, $z_b$ axis in the vehicle plane of symmetry and pointing down perpendicular to $x_b$, $y_b$ axis perpendicular to $x_b$ and $z_b$, pointing to the right wing. 

    \item Propeller Frame $\Gamma_p^i$ : Origin in the center of rotation of the $i-th$ rotor, axis directions aligned with the body frame $\Gamma_b$ when the tilting angles $g_i$ and $b_i$ are zero.
    
    \item Wind Frame $\Gamma_w$: Origin in the airplane COG, $x_w$ axis in the direction of the projection of the wind speed vector in the vehicle plane of symmetry, $y_w$ perpendicular to $x_w$, pointing to the right wing and $z_w$ in the vehicle plane of symmetry and perpendicular to $x_w$ to complete a right handed system, .
\end{itemize}
An overview of the Control, Earth, Body, Propeller and Wind frames and the identification of the rotor tilting angles is shown in Figure \ref{Reference_frames_overview}.

\subsection{Transformation matrices}

For the coordinate transformation between the body reference frame $\Gamma_b$ to control reference frame $\Gamma_c$ the following matrix is used:
\begin{equation}
\scalemath{1}{
    \begin{aligned}
        R_{cb} = \left[\begin{matrix}
        c_\theta & s_\phi s_\theta & c_\phi s_\theta \\
        0 & c_\phi & -s_\phi\\
        -s_\theta & s_\phi c_\theta & c_\phi c_\theta \\
        \end{matrix} \right]&  \\
       \text{such that } \bar{x}_c = R_{cb} \cdot \bar{x}_b \hspace{0.3cm} &
    \end{aligned},
    }
\label{Transformation_matrix_R_cg}
\end{equation}
where $c$ and $s$ represent the abbreviation of the cosine and sine function respectively, while $\phi$, and $\theta$ are the Euler angles in the traditional ZYX order. 

For the coordinate transformation between the earth reference frame $\Gamma_e$ and the control reference frame $\Gamma_c$ the following matrix is used:
\begin{equation}
\scalemath{1}{
    \begin{aligned}
        R_{ec} = \left[\begin{matrix}
        c_\psi & s_\psi & 0 \\
        -s_\psi & c_\psi & 0 \\
        0 & 0 & 1\\
        \end{matrix} \right]&  \\
       \text{such that } \bar{x}_e = R_{ec} \cdot \bar{x}_c \hspace{0.3cm} &
    \end{aligned},
    }
\label{Transformation_matrix_R_bg}
\end{equation} 
where $\psi$ is the yaw Euler angle in the traditional ZYX order.

For the coordinate transformation between the propeller frame $\Gamma_p$ to body reference frame $\Gamma_b$ the following matrix is used:
\begin{equation}
        \begin{aligned}
        R_{bp}^i = \left[\begin{matrix}
        c(b^i) & 0 & s(b^i) \\
        s(g^i) s(b^i) & c(g^i) & -s(g^i) c(b^i) \\
        -c(g^i) s(b^i) & 
        s(g^i) & c(g^i) c(b^i) \\
        \end{matrix} \right] & \\
        \text{such that } \bar{x}_b = R_{bp}^i \cdot \bar{x}_p^i \hspace{.8cm} &
    \end{aligned},
\label{Transformation_matrix_R_pb}
\end{equation}
where the angles $b_i$ and $g_i$ are the i-th rotor tilting angles. Conventionally, within the paper we will refer to $b_i$ as elevation tilting angle and to $g_i$ as azimuth tilting angle. For a visual representation of the tilting angles, the reader can refer to Figure \ref{fig:rotor_angle_definition}. 

Concerning the coordinate transformation between the wind frame $\Gamma_w$ and the body frame $\Gamma_b$, the following matrix is used: 
\begin{equation}
    \begin{aligned}
        R_{bw} = \left[\begin{matrix}
        c(\alpha)c(\beta) & -c(\alpha)s(\beta) & -s(\alpha) \\
        s(\beta) & c(\beta) & 0 \\
        s(\alpha)c(\beta) & -s(\alpha)s(\beta) & c(\alpha) \\
        \end{matrix} \right] & \\
        \text{such that } \bar{x}_b = R_{bw} \cdot \bar{x}_{wb} \hspace{.8cm} &
    \end{aligned},    
\label{Transformation_matrix_R_wb}
\end{equation}
where $\alpha$ is the angle of attack and $\beta$ is the sideslip angle. 

Finally, we define the matrix T to obtain the rate of change of the Euler angles from the body rates $\boldsymbol{\omega}$:
\begin{equation}
    \begin{aligned}
         T = \left[\begin{matrix}
         1  & \text{sin}(\phi)\text{tan}(\theta) &  \text{cos}(\phi)\text{tan}(\theta)   \\
        0   &    \text{cos}(\phi)        &     -\text{sin}(\phi)       \\
        0  & \frac{\text{sin}(\phi)}{\text{cos}(\theta)} &  \frac{\text{cos}(\phi)}{\text{cos}(\theta)}\\
        \end{matrix} \right] & \\
       \text{such that }\dot{\boldsymbol{\Theta}} = T \cdot \boldsymbol{\omega} \hspace{1.3cm} &
    \end{aligned},
    \label{Transformation_matrix_T}
\end{equation}
where $\boldsymbol{\Theta}$ represents the Euler angle vector composed of $\phi$, $\theta$ and $\psi$.
\subsection{Assumptions}
In order to facilitate the EOM derivation, a few assumptions are made:
\begin{itemize}
    \item The inflow angle of the propeller is assumed not to influence its performance.
    \item The thrust generated by the rotor is always perpendicular to the propeller disk and is applied in the center of the propeller disk.
    \item The change in the body inertia due to the rotor tilting is negligible and $x_b$, $y_b$ and $z_b$ are vehicle principal axes.
    \item $x_p$, $y_p$ and $z_p$ are principal axes for the propeller, and the inertia terms $I^p_{xx}$ and $I^p_{yy}$ are negligible.
\end{itemize}

\subsection{Equations Of Motion derivation}
With the reference frames and assumptions defined, it is possible to analyze all the forces and moments contributing to the system dynamics for the development of the EOM: 
\begin{equation}
    \left\{ 
    \scalemath{0.9}{
    \begin{aligned}
    \ddot{P_c} &= \frac{1}{m} \left( F^p + F^a\right) + g R_{ce} \hat{z}_e \\[.3cm]
    \dot{\omega} &= I_b^{-1}\left(- \omega \times I_b \omega + M^t + M^d + M^i + \right.\\
    & \hspace{.3cm} \left. + M^p + M^a + M^{\delta_a} + M^r + M^{tilt}\right) \\[.3cm]
    \end{aligned}
    }
    \right. ,
    \label{Equations_of_motion}
\end{equation} 
where $\ddot{P_c}$ are the linear accelerations in the control reference frame and $\dot{\omega}$ represents the angular acceleration. \\
Each term of equation (\ref{Equations_of_motion}) refers to a specific contribution as follows:
\begin{itemize}
    \item $F^p : $ Forces produced by the propeller thrust projected onto the control frame: \\
    \begin{equation}
        F^p = \sum_{i=1}^{N} R_{cb}  R_{bp}^i \left( 
        \begin{matrix}
        0 \\
        0 \\
        - K_p^T \Omega_i^2 
        \end{matrix} 
        \right),
    \end{equation} \\
    where $K_p^T$ is the thrust coefficient of the motor and $\Omega_i$ is the rotational speed of the i-th motor. 
    \item $F^a : $ Aerodynamic forces produced by the vehicle in the control frame: \\
    \begin{equation}
        F^a = R_{cb}  R_{bw} \left( 
        \begin{matrix}
        -D \\
        Y \\
        -L 
        \end{matrix} 
        \right),
    \end{equation} \\
    where $D$, $Y$ and $L$ are the aerodynamics forces acting on the vehicle and can be expressed as follows \cite{Aircraft_system_identification_book}:
    \begin{equation}
        \left( 
        \scalemath{0.9}{
        \begin{matrix}
        D \\
        Y \\
        L 
        \end{matrix} 
        }
        \right) = Q \left( 
        \scalemath{0.9}{
        \begin{matrix}
        C_{D0} + k_{cd} (C_{L0} + C_{L\alpha}\alpha)^2 \\
         C_{Y\beta} \beta \\
        C_{L0} + C_{L \alpha} \alpha 
        \end{matrix}
        }
        \right),
        \label{Aero_terms_forces}
    \end{equation} 
    with 
    \begin{equation}
        \scalemath{0.9}{
        Q = \frac{1}{2}\rho S V_{a}^2
        },
        \label{Aero_terms_forces_2}
    \end{equation}     
    and where $\rho$ is the air density, $S$ is the wing surface and $V_{a}$ is the airspeed.

    \item $M^a : $ Aerodynamic moments acting on the vehicle in the body reference frame: \\
    \begin{equation}
    \begin{split}
         &\hspace{1.5cm}M^a = \left(
         \scalemath{0.9}{
        \begin{matrix}
        M_L^a \\
        M_M^a \\
        M_N^a 
        \end{matrix} 
        }
        \right) = \\
        = \; &Q\left( 
        \scalemath{0.9}{
        \begin{matrix}
        \bar{b}(C_{M_{L_0}} + C_{M_{L_\beta}}\beta +\frac{\bar{b}}{2V_{tot}}(C_{M_{L_p}}p + C_{M_{L_r}}r)) \\
        \bar{c}(C_{m0}+C_{m\alpha}\alpha)\\
        b(C_{np}\frac{\bar{b}}{2V{tot}}p + C_{nr}\frac{\bar{b}}{2V_{tot}}r)
        \end{matrix} 
        }
        \right)
        \end{split},
        \label{Aero_terms_moments}
    \end{equation} \\
    where $b$ is the wing span and $p$, $q$ and $r$ are the body angular rates that form the vector $\boldsymbol{\omega}$.
    \item $M^{\delta_a} : $ Aerodynamic roll moment acting on the vehicle due to the aileron deflection in the body reference frame: \\
    \begin{equation}
         M^{\delta_a} = \left(
         \scalemath{0.9}{
        \begin{matrix}
        M_L^{\delta_a} \\
        0 \\
        0 
        \end{matrix} 
        }
        \right) = \left( 
        \scalemath{0.9}{
        \begin{matrix}
        Q \; \bar{c} \; C_{M_L}^{\delta_a} \; \delta_a \\
        0\\
        0
        \end{matrix} 
        }
        \right),
        \label{Aileron_term}
    \end{equation} \\
    where $\delta_a$ is the aileron deflection. 
    The aerodynamic coefficients present in Equation \ref{Aero_terms_moments}, in Equation \ref{Aero_terms_forces} and in Equation \ref{Aileron_term} can be identified through test flights, CFD analysis or geometrical vehicle properties \cite{aero_coefficients_book}.
    Within this paper, we will employ the aerodynamic coefficients in Table \ref{tab:aero_coefficients}, estimated though XFLR5 analysis and flight test data.
    Aerodynamic coefficients not mentioned in Table \ref{tab:aero_coefficients} are assumed to be zero.
    
    \item $M^t : $ Torque generated by the rotors due to the propeller thrust: \\
    \begin{equation}
        \begin{aligned}
        \scalemath{0.9}{
        M^t = \sum_{i=1}^{N} 
        }
        \left( R_{bp}^i \cdot 
        \left[ 
        \scalemath{0.9}{
        \begin{matrix}
        0 \\
        0\\
        - K_p^T \Omega_i^2 
        \end{matrix} 
        }
        \right]
        \right)
        \times
        \left( 
        \scalemath{0.9}{
        \begin{matrix}
        l_x^i & l_y^i & l_z^i
        \end{matrix} 
        }
        \right),    
        \end{aligned}
    \end{equation} \\
    
    where $(l_x^i,l_y^i,l_z^i)$ are the coordinates of the i-th rotor in the body reference frame.
    
    \item $M^d : $ Torque generated by the rotors due to the propeller drag: \\
    \begin{equation}
        M^d = \sum_{i=1}^{N}  - R_{bp}^i  
        \left(
        \begin{matrix}
        0 \\
        0\\
        K_p^M \Omega_i^2 
        \end{matrix} 
        \right)
         (-1)^i,
    \end{equation} \\
    where $K_p^M$ is the torque coefficient of the motor.
    \item $M^i : $ Torque generated by the propeller inertia due to the rotational rate change: \\
    \begin{equation}
        M^i = \sum_{i=1}^{N} - J_p  R_{bp}^i 
        \left( 
        \begin{matrix}
        0 \\
        0\\
        \dot{\Omega}_i 
        \end{matrix} 
        \right) 
         (-1)^i,
    \end{equation} \\
    where $J_p$ is the propeller inertia.
    \item $M^p : $ Torque generated by the rotor precession term due to the tilting rotation: \\
    \begin{equation}
       M^p = \sum_{i=1}^{N}   R_{bp}^i  
        \left( 
        \begin{matrix}
        0 \\
        0 \\
        J_p  \Omega_i
        \end{matrix} 
        \right) 
        \times
        \left( 
        \begin{matrix}
        \dot{g}_i \\
        \dot{b}_i \\
        0
        \end{matrix} 
        \right) 
         (-1)^i.
    \end{equation} \\
    
    \item $M^{tilt} : $ Torque generated by the rotor inertial term due to the tilting rotation: \\
    \begin{equation}
       M^{tilt} = \sum_{i=1}^{N}   R_{bp}^i  
        \left( 
        \begin{matrix}
        \ddot{g}_i I_{{xx}_i}^{tilt}\\
        \ddot{b}_i I_{{yy}_i}^{tilt}\\
        0
        \end{matrix} 
        \right) 
         (-1)^i,
    \end{equation} \\
    where $I_{{xx}_i}^{tilt}$ and $I_{{yy}_i}^{tilt}$ are the i-th rotor tilting inertia in the propeller reference frame.
    \item $M^r : $ Inertial term of the rotor due to the vehicle rates \\
    \begin{equation}
        M^r = \sum_{i=1}^{N} - \omega \times \left( R_{bp}^i  \left( 
        \begin{matrix}
        0 \\
        0 \\
        J_p \Omega_i
        \end{matrix} 
        \right)  (-1)^i
        \right).
    \end{equation} 
\end{itemize}

\begin{table}
\centering
\renewcommand{\arraystretch}{1.5}
\begin{tabular}{P{3.5cm}  P{3.5cm} } 
\hline
\multicolumn{2}{c}{\bfseries Aerodynamic coefficients of the vehicle} \\
\hline
$C_{D0}$ & 0.38 \\
\hline
$k_{cd}$ & 0.2 \\
\hline
$C_{L_0}$ & 0 \\
\hline
$C_{L_\alpha}$ & 3 \;$rad^{-1}$\\
\hline
$Cm_{0}$ & 0.05\\
\hline
$Cm_{\alpha}$ & - 0.05 \;$rad^{-1}$\\
\hline
$C_{M_L}^{\delta_a}$ & 0.12 \; $rad^{-1}$\\
\hline
\end{tabular}
\vspace{.1cm}
\caption{Aerodynamic coefficients of the vehicle. The coefficients are estimated through XFLR5 analysis and flight test experiments. The aerodynamic coefficient not mentioned in the table are assumed to be zero.}
\label{tab:aero_coefficients}
\end{table}

\subsection{Simplification of the EOM for the Nonlinear CA algorithm}
To decrease the computational load of the Nonlinear CA method, only the main terms of the EOM in Equation \ref{Equations_of_motion} are considered. The vehicle dynamics $f_s(\boldsymbol{x},\boldsymbol{u})$ considered for the nonlinear optimization process in Equation \ref{Nonlinear_CA_method} is then the following:  
\begin{equation}
    \scalemath{0.95}{
    \begin{cases}
    \ddot{P_c} = \frac{1}{m} \left( F^p + F^a\right) + g  \hat{z}_e \\[.3cm]
    \dot{\omega} = I_b^{-1}\left(- \omega \times I_b \omega + M^t + M^d  + M^a  + M^{\delta_a}\right) \\[.3cm]
    f_s(\boldsymbol{x},\boldsymbol{u}) = ( \ddot{P_c} \;, \; \dot{\omega} )^T \\
    \end{cases}
    }
    \label{Nonlinear_CA_dynamics_simplified}
\end{equation}

\end{document}